\documentclass[%
 aps,
 prl,
 reprint,
 amsmath,
 amssymb,
 showkeys,
 showpacs,
 groupedaddress,
 longbibliography,
 nofootinbib,
 nobibnotes,
 floatfix,
]{revtex4-1}

\usepackage[utf8]{inputenc}
\usepackage{graphicx}
\usepackage{bm}
\usepackage{xcolor}
\usepackage{enumerate}
\usepackage{amsfonts, amsmath, amssymb}
\usepackage[english]{babel}

\usepackage{hyperref}
\hypersetup{
	colorlinks = true,
	linkcolor = black,
	citecolor = black,
	urlcolor = black,
	filecolor = black,
	linktoc = all,
}

\newcommand{\levap}{l_{\textrm{ev}}}
\newcommand{\tevap}{\tau_{\textrm{ev}}}
\newcommand{\tform}{\tau_{f}}
\newcommand{\lpl}{l_{\textrm{pl}}}
\newcommand{\mpl}{m_{\textrm{pl}}}
\newcommand{\tpl}{t_{\textrm{pl}}}
\newcommand{\pp}{\partial}
\newcommand{\ehat}{\hat{e}}
\newcommand{\diag}{\textrm{diag}}
\newcommand{\del}{\nabla}
\newcommand{\plotwidth}{1.5in}
\newcommand{\rbg}{\rho_{\textrm{bg}}}
\newcommand{\rcore}{r_{\textrm{core}}}
\newcommand{\fec}{\chi_{\textsc{fec}}}
\newcommand{\wec}{\chi_{\textsc{wec}}}
\newcommand{\nec}{\chi_{\textsc{nec}}}
\newcommand{\dtr}{\chi_{\textsc{dtr}}}

\newcommand{\evilpad}[3]{\hspace{#1}{#2}\hspace{#3}}

\begin{document}


\title{Understanding black hole evaporation using explicitly computed Penrose diagrams}

\author{Joseph C Schindler}
\email{jcschind@ucsc.edu}
\author{Anthony Aguirre}
\author{Amita Kuttner}
\affiliation{ University of California Santa Cruz, Santa Cruz, CA, USA}

\date{August 2018}

\begin{abstract}
Explicitly computed Penrose diagrams are plotted for a classical model of black hole formation and evaporation, in which black holes form by the accretion of infalling spherical shells of matter and subsequently evaporate by emitting spherical shells of Hawking radiation. This model is based on known semiclassical effects, but is not a full solution of semiclassical gravity. The method allows arbitrary interior metrics of the form $ds^2=-f(r)\,dt^2+f(r)^{-1}\,dr^2+r^2\,d\Omega^2$, including singular and nonsingular models. Matter dynamics are visualized by explicitly plotting proper density in the diagrams, as well as by tracking the location of trapped surfaces and energy condition violations. The most illustrative model accurately approximates the standard time evolution for black hole thermal evaporation; its time dependence and causal structure are analyzed by inspection of the diagram.  The resulting insights contradict some common intuitions and assumptions, and we point out some examples in the literature with assumptions that do not hold up in this more detailed model. Based on the new diagrams, we argue for an improved understanding of the Hawking radiation process, propose an alternate definition of ``black hole" in the presence of evaporation, and suggest some implications regarding information preservation and unitarity.
\end{abstract}

\pacs{04.20.Cv, 04.70.Bw, 04.70.Dy}
\keywords{general relativity, semiclassical gravity, black hole evaporation, Penrose diagram}

\maketitle



\section{I. \ \ Introduction}

The discovery that black holes (BHs) theoretically evaporate~\cite{Hawking:1974rv,Hawking:1974sw,Hawking:1976ra,Hartle:1976tp} in apparently thermal radiation has raised a number of fundamental questions over the years, chief among them the issue of how to reconcile such evaporation with unitary evolution that preserves information, and how to reconcile descriptions of a BH by an infalling observer with those of an exterior observer.

We contend that a full understanding of these issues can greatly benefit from a more detailed understanding of the spacetime structure of an evaporating black hole.  Accordingly, in this work we provide a well-defined class of spacetimes representing a classical model of BH formation and evaporation, and construct explicitly computed Penrose diagrams for these models using the formalism recently developed in~\cite{Schindler:2018wbx}.

Our model corresponds to spherical BHs which form by accreting infalling spherical null shells of matter, and evaporate by emitting spherical null shells of Hawking radiation from near the horizon (notably, the emission location is fixed by energy conservation considerations). Interior metrics including both singular and nonsingular centers are within the scope of our methods. In the most detailed version of the model, continuous time evolution for both the accretion and evaporation processes is approximated by the use of many shells.

The basic structure of the model is motivated by its similarity to renormalized stress tensors usually associated with the evaporation process~\cite{Davies:1976ei}. It is quite similar in spirit to the models first presented by Hiscock~\cite{Hiscock:1980ze,Hiscock:1981xb} and Hayward~\cite{Hayward:2005gi}, and our more sophisticated diagrams may be roughly thought of as numerically calculated versions of the diagram that Hayward originally sketched (although his and our models do differ slightly). By plotting Penrose diagrams for this model in a way that accurately represents both the global and local causal structure of the exact four-dimensional geometry, we are able to attain a more detailed view of the structure of an evaporating BH metric than was previously possible.

Many aspects of the ``true" dynamics of an evaporating BH remain unknown. There are many proposals for the BH end state~\cite{Rovelli:2014cta,Haggard:2014rza,Poisson:1989zz}, and questions have been raised about in what regimes evaporation dominates the dynamics~\cite{Martin-Dussaud:2019wqc}. Nonetheless, BH evaporation appears to be a phenomenon which probably can occur, and for now seems likely to dominate if the environment becomes sufficiently cold. If so, one expects evaporation from macroscopic masses down to near the Planck scale, at which point semiclassical arguments fail and many alternatives seem plausible.

But while the physical relevance of BH evaporation may be up for debate, its importance in the literature certainly is \mbox{not --- its} study has spawned some famous questions~\cite{Hawking:1976ra,Susskind:1993if,Almheiri:2012rt}, and reasoning about these questions almost invariably draws on assumptions about classical spacetime diagrams. 

There is no reason to think that \textit{any} classical model can fully capture a spacetime in which quantum effects are important. But in limits where classical relativity is useful, a more detailed BH model can provide some insight. Here we investigate a simple, concrete model, which hopefully captures most generic aspects of an evaporating BH spacetime, and for which explicit Penrose diagrams can be attained. No direct calculation of the semiclassical stress tensor is attempted, but we point out various reasons to think the model is closer to self-consistency than existing models. A more complete semiclassical treatment may be possible in the future.

We use the new diagrams to address, within the context of our classical model, some persistent questions about BH spacetimes:
\begin{itemize}
    \item Can a BH form in a finite time as viewed by a distant observer? \textit{(Answer: Yes.)}
    \item Of the many types of ``horizon" associated with BH spacetimes (event, apparent, dynamical, trapping, Killing, etc.), which one has physical meaning in terms of the BH boundary? (\textit{Answer: The apparent horizon at the hypersurface $r=2m$.%
    \footnote{This hypersurface is the unique spherically symmetric marginally trapped tube in spacetime, making it a distinguished hypersurface foliated by apparent horizons~\cite{Bengtsson:2010tj}. It is also the boundary of the unique spherical ``trapping nucleus" (see appendix). This boundary is associated with trapped surfaces, and is not a ``causal horizon" in the traditional sense. Its physical significance within the model is established in Sec.~II. Non-spherical generalizations are discussed in Sec.~VIII.
    }%
    )}
    \item What is the causal structure of the apparent horizon? \textit{(Answer: The outer apparent horizon is timelike during evaporation, and spacelike during accretion.)}
    \item Within the context of a purely classical model, does it make sense to think of Hawking radiation as being emitted from a certain location? And if so, where? \textit{(Answer: Yes, from just outside the apparent horizon, which during evaporation is a timelike surface.)}
    \item In evaporating nonsingular BH models, in terms of causal structure: Where exactly are the regions of high density and pressure, where are energy conditions violated, and where are the inner and outer horizons relative to the core? \textit{(Answer: See diagrams.)}
    \item Is it possible to escape the trapped region if the BH is evaporating? \textit{(Answer: Yes, but it requires very good timing. Falling too far past the apparent horizon ensures your destruction in the singularity or (in nonsingular models) core.)}
\end{itemize}
Within our model, these questions have clear, unambiguous answers. To what extent these questions continue to make sense in a fully quantum gravitational description remains unknown.

We also apply these diagrams to the analysis of some broader questions: What is the proper definition of a~BH? What is the correct interpretation of the Hawking radiation mechanism? Is information preserved during BH evolution? The discussion of these and other questions is taken up in the later sections of the paper.

\begin{figure}[t]
     \includegraphics[width=1.4in, trim={0 3mm 0 13mm}, clip]{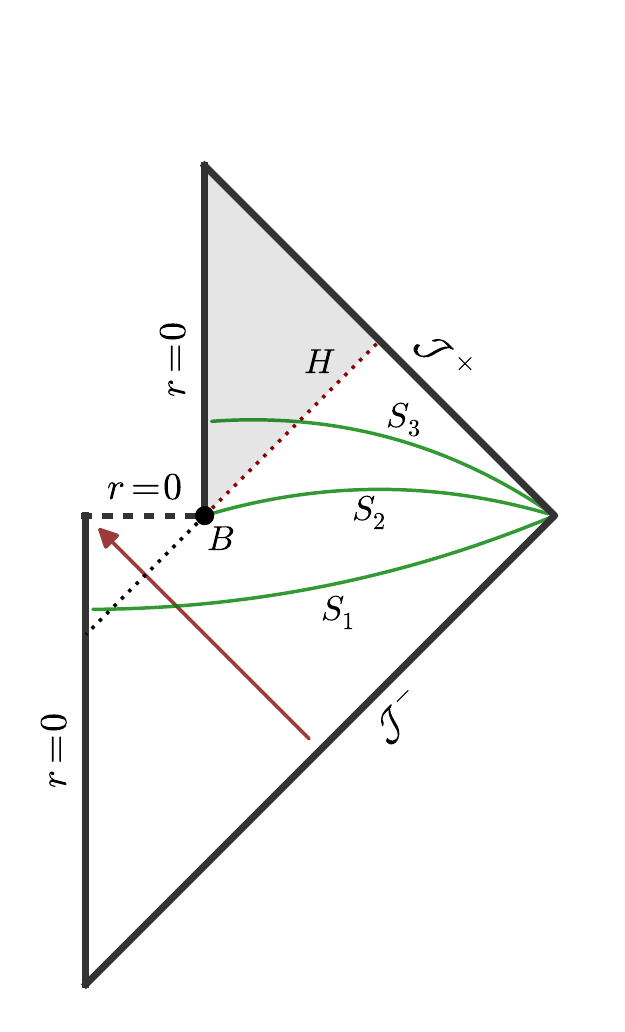}
     \caption{(Color online). Conventional Penrose diagram of an evaporating black hole.  The hole forms via the infall of null dust (red arrow) and forms an event horizon (black dotted).  At some point the black hole evaporates, leaving Minkowski space. Such a diagram usually does not represent any spacetime --- the mechanism of evaporation is left ambiguous, and the nature of point $B$ is totally unclear. In the obvious classical-spacetime interpretations of this diagram, however, point $B$ must be considered a naked singularity that creates a Cauchy horizon $H$ (red dotted), raising questions about the applicability of this diagram for analyzing potentially unitary BH evaporation. Below, we attempt to eliminate these ambiguities by constructing Penrose diagrams for well-defined spacetime models of BH evolution.}
     \label{fig:traditional-diagram}
 \end{figure}

The traditional spacetime diagram for an evaporating BH is depicted in Fig.~\ref{fig:traditional-diagram}. It is essentially the outline of a Schwarzschild BH (formed by collapse) attached to Minkowski space in an unspecified way. There are a few reasons to be wary of this diagram, and to think it may benefit from a more formal treatment.

First, the traditional diagram is ambiguous about what spacetime it is meant to represent --- it does not correspond to any particular model of BH evolution. The mechanism of evaporation is left unspecified, and the nature of point $B$ (where all physics of the evaporation process is hidden) is totally unclear. Since a sketch like this inherently captures no more information than the assumptions put into creating it, it is difficult to learn anything from such a diagram.

Second, any reasonable translation of the traditional diagram to a classical spacetime has a naked singularity and corresponding Cauchy horizon; this can be shown both in simple examples and on general grounds.%
\footnote{In the simplest rigorous interpretation of the standard diagram, where the BH is annihilated by an incoming spherical null shell (for example as in Fig. \ref{fig:simple-schematic}a), a point must be excised from the final Minkowski space for gluing to be topologically allowed near $B$, which creates a Cauchy horizon. The general argument relies on theorems of Geroch \cite{Geroch:1970uw} --- since the domain of dependence of a surface is globally hyperbolic, the assumption that the region beyond $H$ is determined by $S_1$ contradicts continuity of the past and future volumes within a globally hyperbolic space.
} %
This appears to have been noticed at least by Hawking~\cite{Hawking:1974sw} and Birrell and Davies~\cite{Birrell:1982ix} from the beginning, but is forgotten in most modern discussions. In particular, many discussions of the information preservation problem (see for instance~\cite{Susskind:1993if} for a highly referenced example) make essential use of supposed Cauchy surfaces within this diagram, including surfaces on both sides of the evaporation event. But if the spacetime underlying the diagram has a naked singularity and is not globally hyperbolic, no such Cauchy surfaces can be assumed to exist. Because of this hidden assumption of unpredictability, the usefulness of such a diagram in analyzing possibly unitary BH evaporation must be called into question.

Recognizing these shortcomings, and that evaporation may profoundly change the character of the BH spacetime diagram, a number of studies have suggested improved diagrams that more easily allow an interpretation in which information is preserved~\cite{Hayward:2005gi,Ashtekar:2005cj,Horowitz:2009wm,Hossenfelder:2009xq,Hossenfelder:2009fc,Aguirre:2011ac,Stephens:1993an,Nomura:2012cx,Haggard:2014rza,Rovelli:2014cta,Malafarina:2017csn}. These form a useful background for investigating BH evaporation and related issues, and we build most directly on the work of Hayward~\cite{Hayward:2005gi}, who has provided the most minimal and generic model.

By extending this type of model to a form in which explicit Penrose diagrams can be attained, we explore the structure of these improved models while resolving the ambiguity and hidden assumptions inherent to hand drawn diagrams. The new diagrams (Figs. \ref{fig:simple-models}--\ref{fig:background-curvature}) are simultaneously both Penrose diagrams and exact coordinate diagrams, allowing a detailed picture of the exact geometry. A discussion of the diagram formalism, and an explanation of some key aspects of interpreting the new diagrams, is provided in Section~IV.

 \section{II. \ \ Shell model of black hole\\ formation and evaporation}
 \label{sec:model}

 We model the process of black hole formation and evaporation according to the following assumptions:
\begin{enumerate}[(i)]
    \item The black hole is non-rotating and spherically symmetric.
    \item The process is quasistatic, allowing dynamical evolution to be modeled by a sequence of equilibrium BH solutions joined across null shells of matter (such null shells may represent either truly light-like radiation, or highly accelerated timelike matter).
    \item The equilibrium black hole solutions locally have the form $ds^2=-f(r)\,dt^2+f(r)^{-1}\,dr^2+r^2\,d\Omega^2$.
    \item Stellar collapse and mass accretion is modeled by a sequence of ingoing spherical null matter shells, incident from infinity.
    \item Hawking radiation is modeled by pairs of spherical null matter shells. Each pair consists of an outgoing positive-mass shell and ingoing negative-mass shell. Each pair nucleates at a fixed radial distance $\levap$ outside the apparent horizon at $r=2m$, with both shells propagating toward the future. Nucleation points violate the DTR relation (an equation related to energy conservation, see appendix), but the amount of violation is arbitrarily small in the $\levap \to 0$ limit.~If $\levap \approx \lpl$, tiny DTR violations may be considered small quantum fluctuations. In this sense, in our semiclassical model, energy conservation forces Hawking radiation to be emitted from just outside the horizon.
\end{enumerate}
This model is a slightly generalized discrete approximation to that proposed originally by Hayward~\cite{Hayward:2005gi}, and the evaporation mechanism agrees, heuristically, with the classic calculation by Davies, Fulling, and Unruh of the stress tensor for a quantum scalar field in the presence of a static BH~\cite{Davies:1976ei}. We construct spacetimes applying this model, and their corresponding Penrose diagrams, by the methods of \cite{Schindler:2018wbx}. It is assumed that physically realistic models are achieved by first taking the limit $\levap \to \lpl$ at each shell of Hawking radiation, then taking the continuous (many-shell) limit.

 \begin{figure}[t]
    \begin{tabular}{|c|c|}
      \hline
      \hfill (a) & \hfill (b)
      \\
      \includegraphics[width=1.4in]{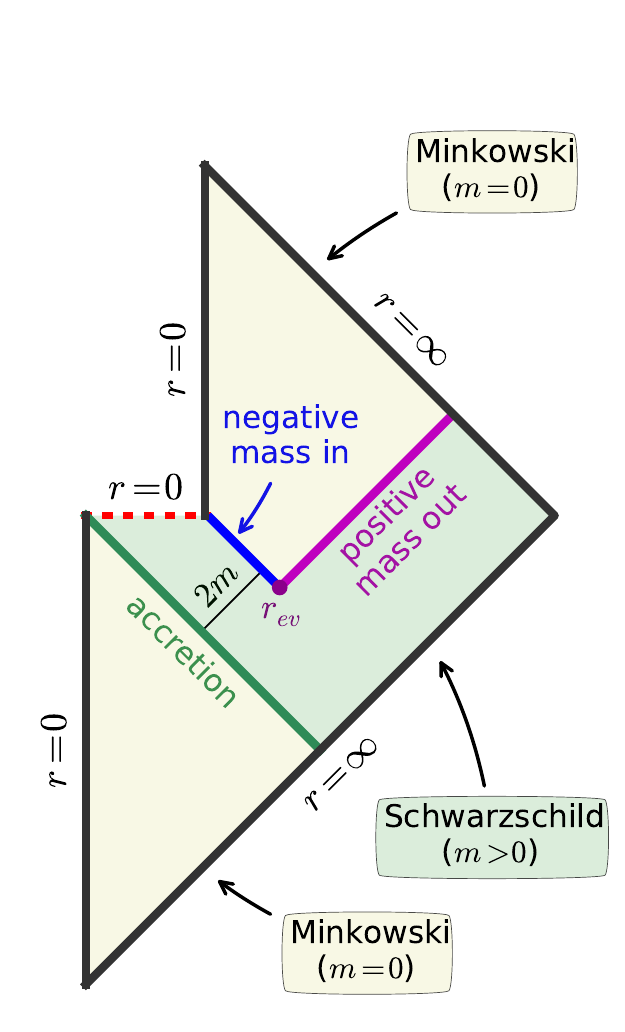}
      &
      \includegraphics[width=1.4in]{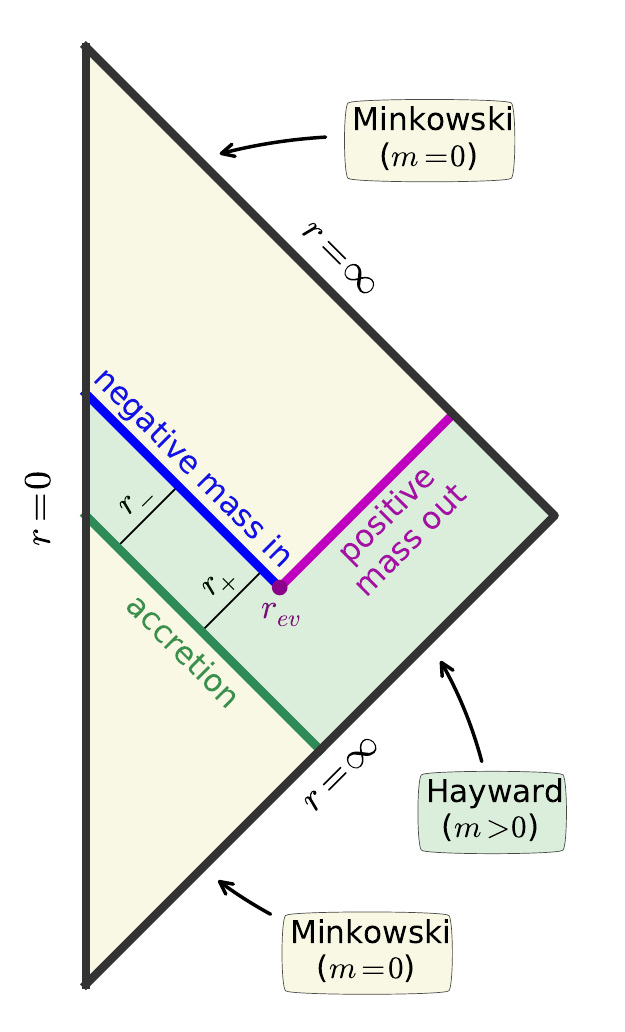}
      \\[2pt] \hline
  \end{tabular}
     \caption{(Color online). Schematic illustration of Penrose diagrams for the shell model in simple cases. (a) The simplest singular case: a Schwarzschild black hole forms by collapse of a spherical null shell, and evaporates by emitting a single burst of Hawking radiation, which nucleates at a radius $r_{\rm ev}=r_{\rm hor}+\levap$ just outside the apparent horizon at $r_{\rm hor}=2m$. (b) The simplest nonsingular case: a Hayward black hole (see below) forms and evaporates in the same way.}
     \label{fig:simple-schematic}
 \end{figure}

 The simplest example of this approach, in which formation and evaporation each occur in a single burst, is depicted schematically in Fig.~\ref{fig:simple-schematic} for both singular and nonsingular interior cases; the exact diagrams will be given later. More realistic models are obtained by using an arbitrarily large number of shells and piecewise regions to approximate the desired smooth dynamics.

 \section{III. \ \ Matter Content of Shell Models with Schwarzschild or Hayward Interior}

 A benefit of explicit diagrams is that matter dynamics during the formation and evaporation processes can be quantitatively tracked. We are concerned with four quantities:
 \begin{enumerate}[(i)]
    \item The proper density and pressures of the bulk (equilibrium) spacetime, defined (up to a sign) as eigenvalues of~${G^\mu}_\nu/(8\pi)$. For metrics of form (\ref{eqn:sss-metric-0}) these include a density $\rho$, a transverse pressure $p_t=-\rho$, and an angular pressure $p_{\Omega}$. Also useful is the mass function $m(r)$ defined by  $f(r)=1-2m(r)/r$.
    \item The location of trapped surfaces in the bulk spacetime, as characterized by the trapped spheres region (see appendix). Trapped spheres occur where $f(r)<0$ in the metric (\ref{eqn:sss-metric-0}), and the trapped spheres region is bounded by apparent horizons where $f(r)=0$ (see appendix).
    \item The location and magnitude of energy condition (EC) violations in both the bulk spacetime and on the null shells, quantified by the EC violation functions $\nec$, $\wec$, and $\fec$ (see appendix).
    \item The local surface density $\sigma$ of null shells, which is proportional to the mass jump $[m(r)]$ across the shell (see appendix). The proportionality is positive (negative) for a shell which is radially ingoing (outgoing) towards the future.
\end{enumerate}
A detailed general analysis of the matter content for models of the present type is given in the appendix.

 Our setup allows for a variety of models of the black hole interior; any metric of the ``strongly spherically symmetric" form
 \begin{equation}
 \label{eqn:sss-metric-0}
     ds^2=-f(r)\,dt^2+f(r)^{-1}\,dr^2+r^2\,d\Omega^2
 \end{equation}
 is allowed. Metrics of this type can be either singular or nonsingular at $r=0$  (for definition and properties see appendix). Nonsingular models have the advantage that all matter is made explicit in the stress tensor, whereas singular solutions contain a matter contribution hidden in the singularity. Although classical theorems do predict singularity formation in gravitational collapse \cite{Hawking:1973uf}, nonsingular solutions are thought to arise in effective semiclassical approximations if quantum gravitational effects regulate curvature at the Planck scale. Nonsingular solutions often violate classical energy conditions, but since quantum field theories are well known to do the same, this is not a major defect~\cite{Martin-Moruno:2017exc}.

 A number of common metrics are of the form (\ref{eqn:sss-metric-0}), including
 \begin{equation*}
     \begin{array}{lcl}
        \textrm{Minkowski}   & \quad & f(r)=1, \\
        \textrm{de Sitter}   & \quad & f(r)=1 - r^2/L^2, \\
        \textrm{Anti de Sitter}   & \quad & f(r)=1 + r^2/L^2, \\
        \textrm{Schwarzschild}   & \quad & f(r)=1 - 2m/r, \\
        \textrm{Reissner-Nordstrom}   & \quad & f(r)=1 - 2m/r + Q^2/r^2, \\
        \textrm{Hayward}   & \quad & f(r)=1 - 2mr^2/(r^3+2ml^2), 
     \end{array}
 \end{equation*}
 among others. We limit for now our consideration to two simple cases: Schwarzschild (singular) and Hayward (nonsingular) interiors.

 \begin{figure}[t]
 \includegraphics[width=2.6in]{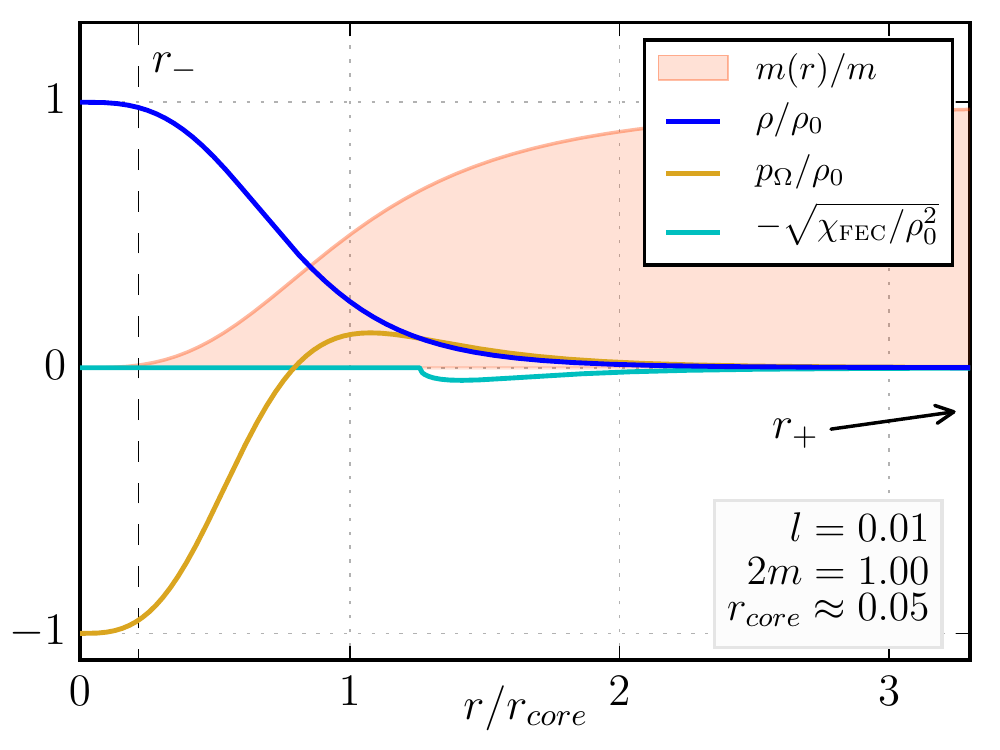}
 \caption{
 \label{fig:hayward-matter}
 (Color online).
 Matter content of the Hayward spacetime. 
 Density and pressure are maximized at $r=0$, with maximum density 
 \mbox{$\rho_0 = \frac{3}{8\pi l^2} = m / \!\left(\frac{4}{3}\pi r_{core}^3\right)$}. The FEC violation function $\fec$ (see appendix) shows non-negligible violations occur near the core surface; \textsc{NEC} and \textsc{WEC} are violated at negative-mass shells in the dynamic model, but not here in the bulk. The plot shows a small BH, but is mostly parameter-independent: increasing $m$ pushes the horizons at $r_{-}\approx l$ (dashed) left and at $r_{+}\approx 2m$ (off scale) right, with no other effects. Spheres are trapped surfaces for all \mbox{$r_{-} < r < r_{+}$}, which we call the trapped spheres region.
 }
 \end{figure}
 
 We take the Schwarzschild metric to model BHs containing a singularity. As is well known, the matter content is trivial: the spacetime contains a vacuum everywhere outside the singularity at $r=0$. Trapped spheres occur in the interval $0<r<2m$, with a horizon located at $r=2m$. For two Schwarzschild solutions of mass parameter \mbox{$m_\pm=m\pm \Delta m/2$} joined at a null shell, the mass jump $[m(r)] = \Delta m$ is a constant, and ingoing (outgoing) positive-mass shells increase (decrease) mass toward the future. There are no EC violations.

 Black holes with nonsingular interior we model by the Hayward metric, with
 \begin{equation}
 \label{eqn:hayward-metric}
     f(r) = 1 - \frac{2m r^2}{r^3 + 2m l^2}.
 \end{equation}
 Density, pressures, and energy condition violations associated with this metric are shown in \mbox{Fig. \ref{fig:hayward-matter}}.

 This metric has parameters $l$ and $m$; $l$ determines the proper density measured at the core and defines the length scale on which curvature is regulated by quantum gravitational effects, while $m$ determines the black hole mass measured by a distant observer. In our models, $l$~is held fixed at a small value, while only $m$ varies across shells. Physically, one assumes that $l \approx \lpl$, and that $2m \gg l$ except in the final moments of evaporation.

 When $2m \gg l$, a radial slice of Hayward spacetime is split into three intervals by horizons at 
 \begin{equation*}
     \begin{array}{lcl}
        r_- \approx l   & \quad & \textrm{(inner horizon),} \\
        r_+ \approx 2m  & \quad & \textrm{(outer horizon),}
     \end{array}
 \end{equation*}
  with trapped spheres occuring in the interval
  \begin{equation*}
     \begin{array}{lcl}
        r_- < r < r_+   & \quad & \textrm{(trapped spheres region).}
     \end{array}
 \end{equation*}

 The matter distribution describes an extremely dense (Planck scale density) core of length scale
 \begin{equation}
        \rcore=(2ml^2)^{1/3},
 \end{equation}
 which (except in the final moments of evaporation) obeys $l \ll \rcore \ll 2m$. It is thus useful to think of the following regions:
  \begin{equation*}
     \begin{array}{lcl}
       r \ll \rcore     ,  & \quad & \textrm{Quantum Gravity Core,} \\
       r \approx \rcore     ,  & \quad & \textrm{Core Surface,} \\
       r \gg \rcore          ,  & \quad & \textrm{Nearly Schwarzschild Vacuum.} \\
     \end{array}
 \end{equation*}
 The core contains a homogeneous distribution of extreme density and pressure (Ricci curvature $ \sim l^{-2}$), with metric closely approximating de Sitter. Far away from the core, the metric closely approximates the traditional Schwarzschild vacuum (Ricci curvature vanishes like $(r/r_{core})^{-6}$). The core surface is characterized by a rapid change in density and pressure accompanied by EC violations. It is satisfying that the core extends outside the inner apparent horizon: despite the lack of a singularity, trapped matter is doomed to quantum gravity decomposition.
 
 When Hayward regions of mass parameter \mbox{$m_\pm=m\pm \Delta m/2$} are joined across a null shell junction, the resulting mass jump is
  {\vspace{-1mm}%
  \begin{equation}
     [m(r)] = \Delta m \; 
     \frac{ \big(\frac{r^3}{2 |\Delta m| l^2} \big)^{\!2} }
     {\big(\frac{r^3}{2 |\Delta m| l^2} + \frac{m}{|\Delta m|}\big)^{\!2} - 1} \;,
  \end{equation}}%
 which develops monotonically from zero at $r=0$ to \mbox{$\Delta m$ at $r\to \infty$} for small $\Delta m/m$. Far from $r=0$, this closely approximates the Schwarzschild case, with a constant mass jump $\Delta m$. Near $r=0$, the shell mass gradually approaches zero over lengths of order $(2 |\Delta m| l^2)^{1/3}$, as the shell is absorbed into (or generated by) the quantum gravity region.

 It is often convenient to describe spacetimes of the form~(\ref{eqn:sss-metric-0}) in double-null Eddington-Finklestein coordinates $(u,v)$ defined by
 \begin{equation}
     \begin{array}{c}
        du = dt - f(r)^{-1} \, dr , \\
        dv = dt + f(r)^{-1} \, dr .
     \end{array}
 \end{equation}
 An integration constant in each coordinate acts as an unphysical overall time translation. For asymptotically flat cases, the coordinate $v$ runs along past null infinity (at constant $u=-\infty)$, while $u$ runs along future null infinity (at constant $v=\infty$).
 
 How generic is the Hayward metric for describing nonsingular BHs? If we restrict to the form (\ref{eqn:sss-metric-0}), very generic: assuming the topology of a stellar-collapse BH, so that $r\to 0$ is included in the spacetime, the asymptotic behaviors as $r\to 0$ and $r \to \infty$ are fixed by physical considerations (nonsingularity, monotonic density, approximately Schwarzschild), so the only freedom in $f(r)$ involves the transition to vacuum at the core surface. Since details of the mass profile at the core surface have no important effect on causal structure (no additional horizons are introduced without a drastic change), the exact form of $f(r)$ is not important. On the other hand, there does exists a freedom to generalize~(\ref{eqn:sss-metric-0})~by including a redshift factor $\alpha(r)$ such that
 \begin{equation}
 \label{eqn:redshift}
     ds^2=- \alpha(r)^2 f(r)\,dt^2+f(r)^{-1}\,dr^2+r^2\,d\Omega^2 ,
 \end{equation}
 while maintaining the same qualitative picture. Including this redshift factor leaves the proper density unchanged, but alters the curvature scalars and proper pressures, in addition to modifying proper times measured by fixed-radius observers. The possibility of including a redshift factor seems physically admissible (so long as care is taken to make sure it introduces no undesirable effects), and is worth further consideration; one interesting application of this approach was given recently in \cite{DeLorenzo:2014pta}. While we do not include the metric (\ref{eqn:redshift}) in our full analysis (it~is outside the scope of our algorithm for generating Penrose diagrams), we do not expect the omission to have a major effect on the resulting~diagrams, since the redshift factor does not significantly alter the causal structure.

In what follows, the matter content that has been described throughout this section will be plotted in spacetime diagrams to visualize the flow of matter during BH formation and evaporation.
 
 \section{IV. \ \ Diagram formalism}

The new diagrams presented here (Figs.~\ref{fig:simple-models}--\ref{fig:background-curvature}), which are constructed using the methods developed in~\cite{Schindler:2018wbx}, may look somewhat strange to those used to only outlines and sketches. A few comments are in order.

The diagrams are obtained by directly finding a global, compact, double-null coordinate system for the spacetime --- in this sense they are not just Penrose diagrams, but also exact spacetime coordinate diagrams.%
\footnote{The usefulness of the direct coordinate approach to causal diagrams was always made clear by Carter~\cite{Carter:1966zza,CARTER1966423} and others (e.g.~\cite{Kruskal:1959vx}), but is not always made explicit in modern treatments, which sometimes put more emphasis on conformal mappings following Penrose's original method~\cite{Penrose:1964ge,Penrose:1965am}. But the conformal transformation aspect of Penrose diagrams is actually slightly misleading in four dimensions, since it is common to construct diagrams for spacetimes which are neither conformally flat nor conformally related to anything interesting (see~\cite{Chrusciel:2012gz}). It is only the two-dimensional diagram plane (normal to the symmetry directions) which is necessarily conformally flat. The coordinate diagram approach is more in the spirit of Carter than Penrose, and it would be justified to alternately call these \mbox{Penrose-Carter} diagrams, but that name is longer and less traditional. For more details about the theory of these diagrams, see~\cite{Schindler:2018wbx}.
\label{footnote:conformal}} %
This allows all aspects of the exact four-dimensional geometry to be captured in the diagram, including both the global and local (interior) causal structure. No conformal information is thrown away. Each point in the diagram represents a spherical symmetry surface, and, because of the symmetry, it makes sense to discuss the exact Riemann curvature (as well as anything derived from it, like proper densities and pressures) and exact geometry at any point of the diagram. We make use of this to precisely plot the matter content, trapped surfaces, and other features within the diagram. That these coordinate diagrams simultaneously act as Penrose diagrams, correctly portraying the causal structure in the usual way, is ensured within our formalism~\cite{Schindler:2018wbx}.

Given any spacetime, there exists the freedom to deform it by arbitrary conformal transformations without disrupting the causal structure. Consequently, there is sometimes assumed to be a corresponding freedom to conformally distort Penrose diagrams. But while it is true that conformal deformations preserve causal structure of the diagram, such deformations do not preserve the geometry of the spacetime which is supposed to be represented. Since we are interested in exactly representing the full spacetime geometry (and not merely conformally related spacetimes), arbitrary conformal distortions to the diagram cannot be allowed. 

Despite this restriction, there remains a large amount of freedom to distort the diagram by change of coordinates. In particular, any change of coordinates which acts conformally on the metric%
\footnote{That is, when restricted to the two-dimensional diagram plane (see footnote~\ref{footnote:conformal}) it alters the metric only by an explicit conformal factor. This preserves the causal structure of the coordinate system~\cite{Schindler:2018wbx}.
} %
yields another valid diagram. In practice, this freedom amounts to separately deforming the $U$ and $V$ coordinates (which are some null coordinates defining the diagram) by any monotonic functions.

While the freedom to deform Penrose diagrams is well known, it is not widely recognized just how different a set of valid deformations can make a diagram appear. We will see in Fig.~\ref{fig:nonsingular-realistic} that three causally equivalent diagrams paint what, at first glance, appear to be three very different pictures of the same spacetime. The key point in reconciling the apparent difference is that in any single diagram, some features are squished beyond recognition. This is unavoidable, since evaporating BH spacetimes contain a number of length/time scales (Planck scale, horizon scale, formation timescale, evaporation timescale, and in nonsingular models: core scale) which can be drastically different. 

For instance, in any model where the BH has a macroscopic mass, the Planck scale, horizon scale, and evaporation timescale (and in nonsingular models, the core size) are all extremely different. Any individual diagram will only clearly represent one of these scales at a time (the rest being squished into points lines and edges). Gaining a clear understanding of the complete causal structure therefore requires the use of multiple diagrams. By carefully inspecting and comparing a few, the full story can be pieced together.

While allowing more general (non-coordinate) conformal transformations in the diagram plane would, in a strictly pointwise-causal-structure sense, allow more scales to be depicted simultaneously by ignoring distance information, it would also (as discussed above) destroy the exact representation of the four-dimensional geometry which is essential to our \mbox{analysis}. This implies a lesson about sketching diagrams: If one wishes to accurately depict details of the internal structure, the class of allowed diagram deformations may be more restricted than naively assumed.

For more details on how these diagrams are constructed, a more general formalism for the analysis of Penrose diagrams (including, for example, a strict definition of what a Penrose diagram is), and further discussion of the relation between Penrose diagrams and conformal transformations, consult~\cite{Schindler:2018wbx}%
\footnote{%
The methods of \cite{Schindler:2018wbx} have been implemented in a software module \texttt{xhorizon} for Python, which is under development by the authors and available under a free open source license at \href{https://github.com/jcschindler01/xhorizon}{https://github.com/jcschindler01/xhorizon}.
}%
.

 \section{V. \ \ Diagrams for simple models}

\begin{figure*}[p]
  \begin{tabular}{ccc}
      \includegraphics[]{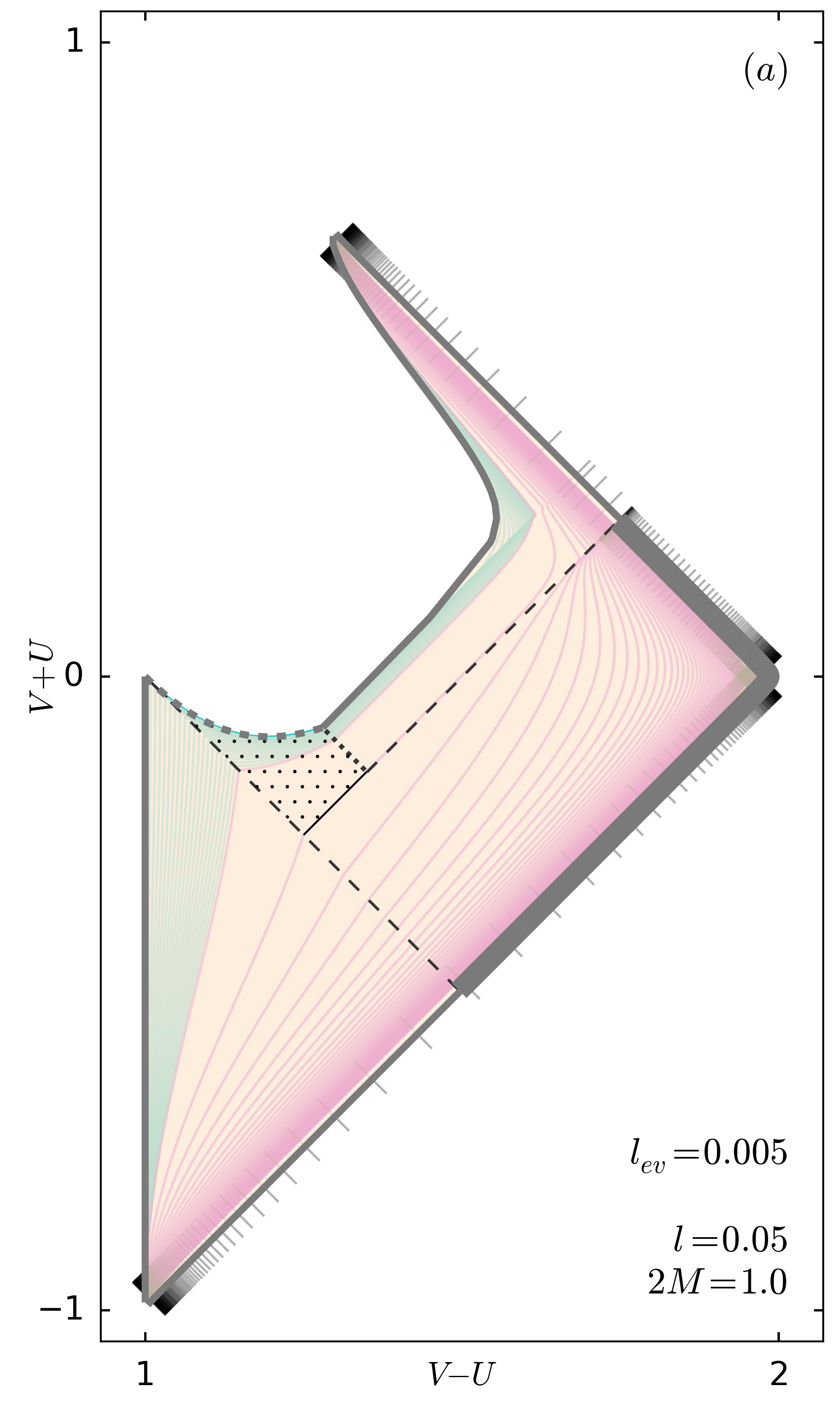}
      &
      \includegraphics[]{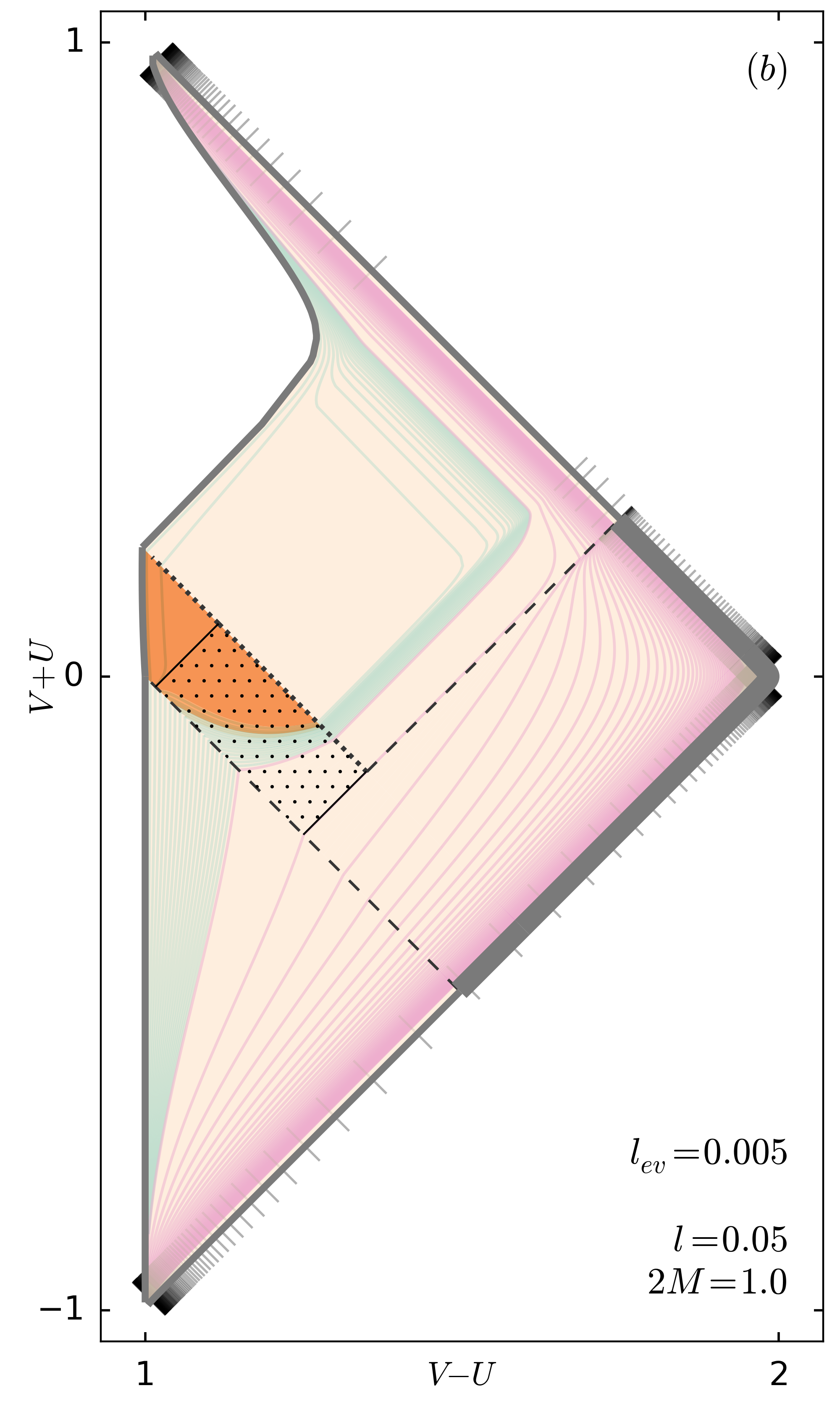}
      &
      \includegraphics[]{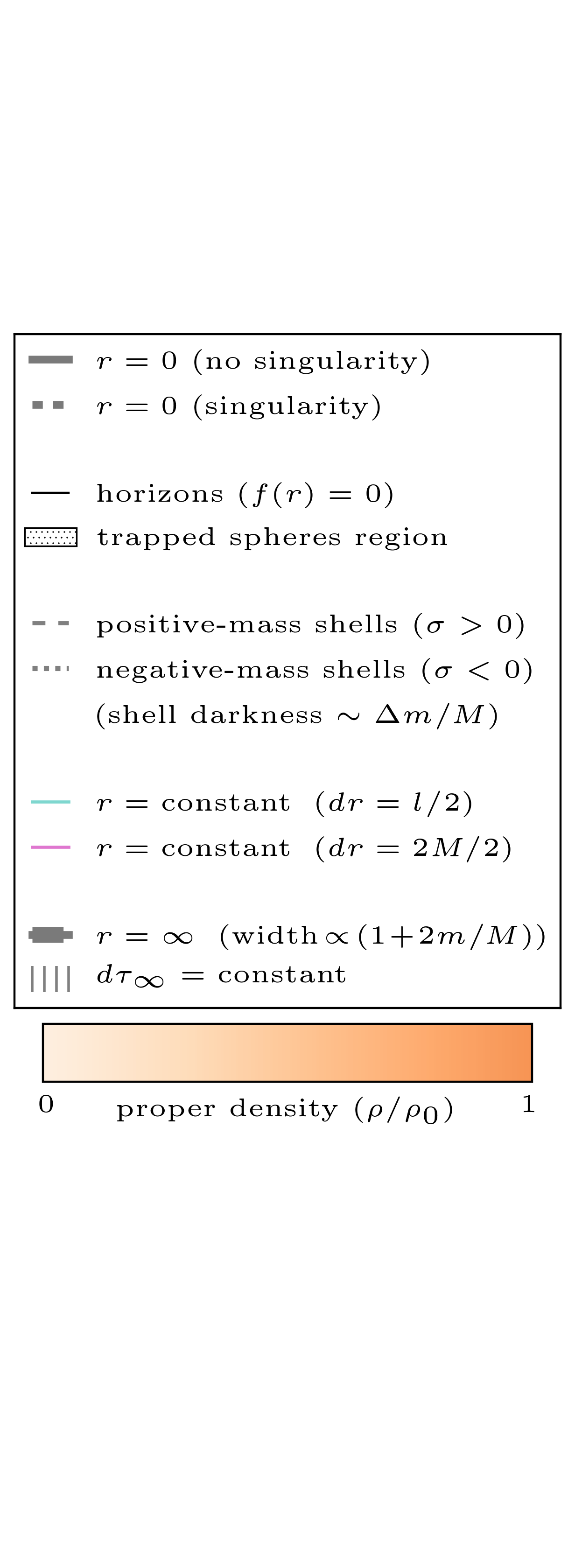}
  \end{tabular}
     \caption{ 
     (Color online).
     Penrose diagrams for (a) singular and (b) nonsingular black holes which form by accreting a single shell of infalling matter and evaporate by emitting a single blast of Hawking radiation (see Fig. \ref{fig:simple-schematic} for an illustrative schematic).
     Parameters are chosen to illustrate qualitative features, but the time evolution and relative length scales are not realistic (see Fig.~\ref{fig:nonsingular-realistic} for improved model).
     Positive-mass (accretion and outgoing Hawking radiation, gray dashed) and negative-mass (ingoing Hawking radiation, gray dotted) shells separate the spacetime into piecewise regions, with Hawking radiation nucleating at a tiny radial distance $l_{ev}$ outside the horizon of the region to its past. In diagrams with many shells, shell masses are indicated by grayscale darkness (darkness proportional to $1+2\,\Delta m/M$). 
     The curvature cutoff length scale $l$ (which has physical significance only in Hayward regions) is held fixed across all regions, while the mass parameter $m$ (which in every region determines the gravitational mass measured by a distant observer) varies. 
     The total mass $M$ is the maximum value of $m$ in any region, and locally $m$ is visualized by the linewidth of the conformal boundary at $r=\infty$ in each region (linewidth proportional to $1+2m/M$). 
     Tick marks (gray) along $r=\infty$ mark off equal increments of proper time for an infinitely distant observer at constant radius (i.e. constant increments of $du$ and $dv$ along null infinity).
     The trapped spheres region (black dot-hatch fill), bounded by horizons (black) where $f(r)=0$, contains closed trapped spheres.
     Background coloring is determined by the local proper density $\rho$ (orange color scale) scaled by the maximum density $\rho_0 = 3/(8\pi l^2)$. 
     The Hayward core is clearly visible as a dark orange region in the density plot, and the core surface almost exactly corresponds to the singularity location in the Schwarzschild case.
     Notably, distant observers near future null infinity begin to observe Hawking radiation at the same moment they see the infalling accretion shell fall through its own horizon.
     Lines of constant radius are shown at small ($dr=l/2$, teal) and large ($dr=2M/2$, magenta) length scales; even where they appear bundled or strongly kinked, they do in fact remain continuous.
     One strange-looking feature of this diagram is the appearance of a set of wiggly kinks and a few stray tick marks to the future (measured along future infinity) of the final evaporation shell, before the very stretched out area; this is an artifact of the unrealistic parameters, and in more realistic models these kinks and tick marks all coincide with the final shell.
     Coordinates $V$ and $U$ defining the axes are basically arbitrary null global coordinates, defined further in \cite{Schindler:2018wbx}.
     The same visualization scheme described here is used in all examples~below.
     }
     \label{fig:simple-models}
 \end{figure*}

 Explicitly computed diagrams are shown in Fig.~\ref{fig:simple-models} for the simple models depicted schematically in Fig.~\ref{fig:simple-schematic}. The parameters were chosen to emphasize certain qualitative aspects, but are not particularly realistic (see Fig.~\ref{fig:nonsingular-realistic} for an improved model). The simple models capture most qualitative features of the more detailed diagrams, and several features in particular are worth noting.

 Immediately apparent is the similar location of the Schwarzschild singularity and the Hayward core surface (this coincidence is generic, see~\cite{Schindler:2018wbx}); the two surfaces almost exactly coincide. The intersection of the singularity with timelike $r=0$ corresponds to the point $B$ in Fig.~\ref{fig:traditional-diagram}, about which questions were raised in the introduction. From a technical perspective, this point creates a Cauchy horizon in Fig.~\ref{fig:simple-models}a since topological matching conditions require that $B$ be excised from the final Minkowski space. It is probably more physical, however, to examine the nature of point $B$ by looking at the corresponding point in Fig.~\ref{fig:simple-models}b. Clearly, a surface like $S_2$ of Fig.~\ref{fig:traditional-diagram}, terminating at the point corresponding to $B$, is not a Cauchy surface in the regularized spacetime. At least to the extent that the singularity cuts off an unknown semiclassical spacetime, this shows on physical grounds that $B$ should be considered a naked singularity. This picture is likely generic to nonsingular extensions of the Schwarzschild solution, as the causal structure of the core does not depend on details of the metric. While the spacetime of Fig.~\ref{fig:simple-models}b is globally hyperbolic, the above analysis has an important consequence: the spacetime of Fig.~\ref{fig:simple-models}a is not.
 
 A natural question to ask about an evaporating BH is whether it's possible to escape after falling through the apparent horizon. Formally, \textit{do there exist timelike curves intersecting the trapped region but avoiding the core or singularity?} According to our shell model, the answer is yes, as is evident from Fig.~\ref{fig:simple-models}. Indeed, as seen even more clearly in Fig.~\ref{fig:nonsingular-realistic}, during evaporation the apparent horizon forms a timelike surface of decreasing radius, despite being locally null within each piecewise region. For an observer who barely crosses the horizon, it's possible to wait for the horizon to ``evaporate past," allowing an escape. Nonetheless, all timelike observers in the trapped region are radially infalling, and will eventually be doomed to destruction in the core if the black hole is long-lived relative to their ability to accelerate.

 Tick marks along null infinity in Fig.~\ref{fig:simple-models} represent constant intervals $du=const$ and $dv=const$ along future and past null infinity respectively. Aside from specifying equal increments of proper time for distant observers, these ticks are useful for analyzing particle creation by the spacetime in quantum field theoretic computations. In the standard analysis, in and out vacuum modes are associated with the $u$ and $v$ coordinates, and the thermal Hawking flux is associated with an infinite phase buildup found by comparing $dv$ increments to traced-back $du$ increments at past null infinity \cite{Hawking:1974sw}. Noting that the tick marks have an additive opacity (so that darker marks actually show many superimposed ticks), it's clear that some approximation to the usual phase buildup effect is present in both Fig.~\ref{fig:simple-models} and the subsequent examples.
 
 While the simple single-burst models capture many aspects of the more realistic diagrams, they also differ in certain respects. Most importantly, realistic models have four very different length scales, corresponding to $\lpl$, $\rcore$, $M$, and the BH lifetime $\tevap$ (where $l$ and $\levap$ are both on the $\lpl$ scale). As a consequence, more realistic diagrams tend to have very sharply kinked features, and features can be clearly portrayed only for one length scale at a time, with other features relegated to lines, bundles, and corners. Another difference between Fig.~\ref{fig:simple-models} and more advanced models is quite noticeable: as also described in the caption, the separation (measured along future infinity) between kinks in the lines $r=const$ and the final evaporation shell is an artifact of the unrealistic parameters of Fig.~\ref{fig:simple-models}. In Fig.~\ref{fig:nonsingular-realistic} and other realistic models, all diagram features associated with the final step of evaporation coincide with the final shell.

 \section{VI. \ \ Diagrams for a nonsingular model with more realistic time evolution}
 \label{sec:nonsingular-realistic}
 
 A more detailed model is attained by approximating continuous time evolution with a large number of shells and piecewise regions. A diagram of this type, constructed based on a Hayward interior of fixed curvature cutoff length scale $l$, is presented in Fig.~\ref{fig:nonsingular-realistic}. Its parameters, justification, and implications are discussed below. While we have chosen here to work with a nonsingular model, all the diagrams can be translated to the Schwarzschild case by simply replacing the core surface with a singularity and ignoring the interior region (see~\cite{Schindler:2018wbx} for why this is valid).

 \renewcommand{\plotwidth}{2.14in}
 \begin{figure*}[p]
 \begin{tabular}{ccc}
      \includegraphics[width=\plotwidth]{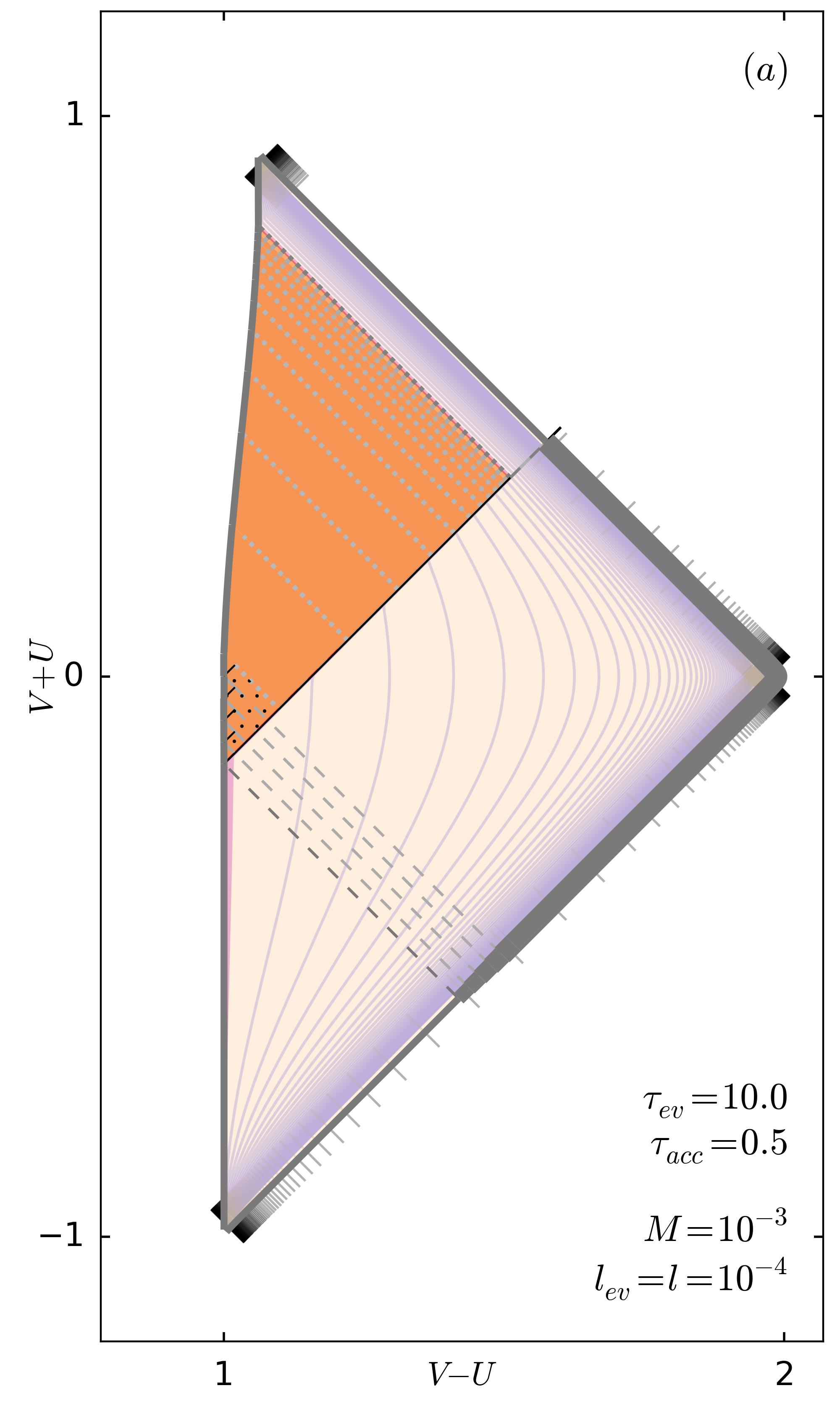}
      &
      \includegraphics[width=\plotwidth]{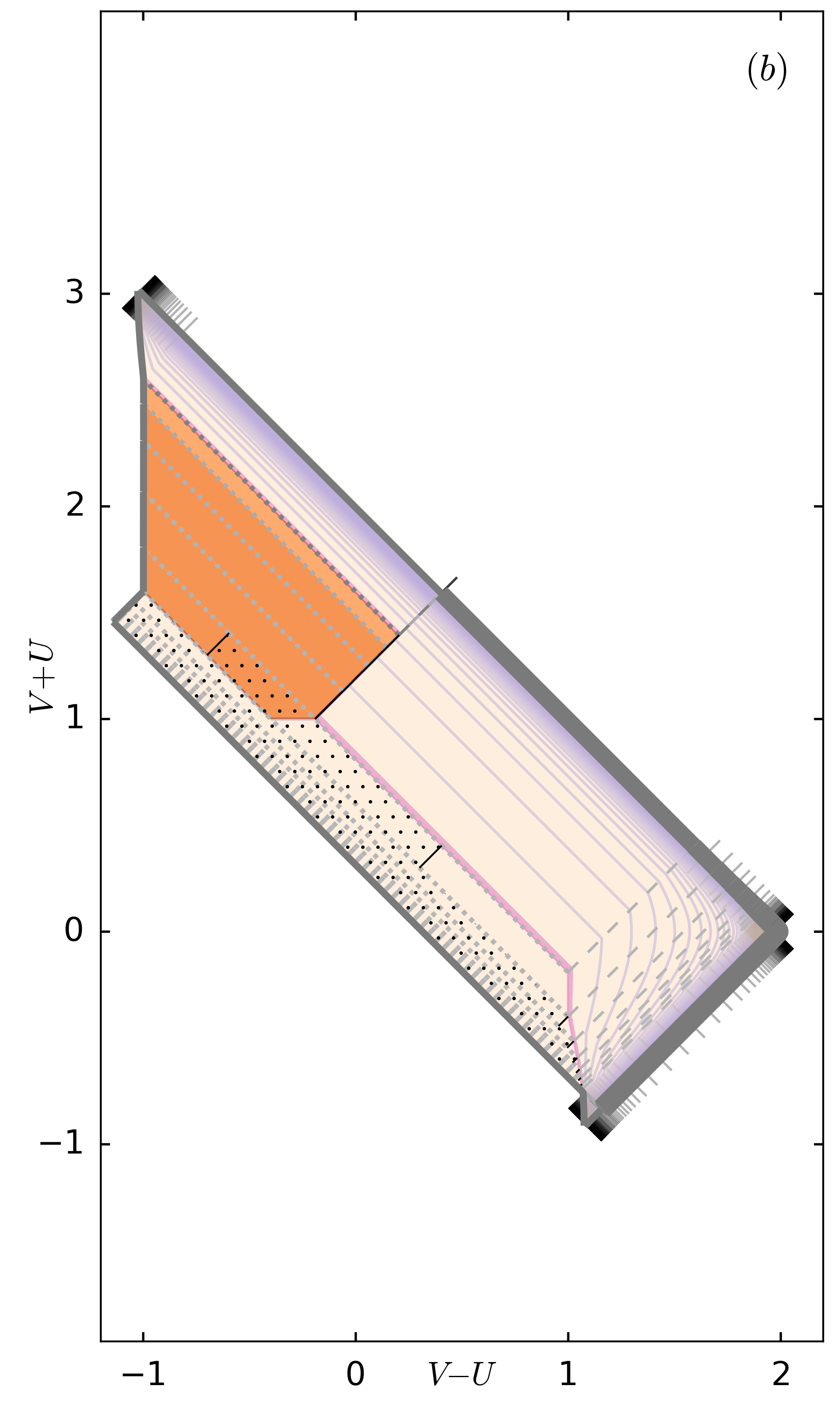}
      &
      \includegraphics[width=\plotwidth]{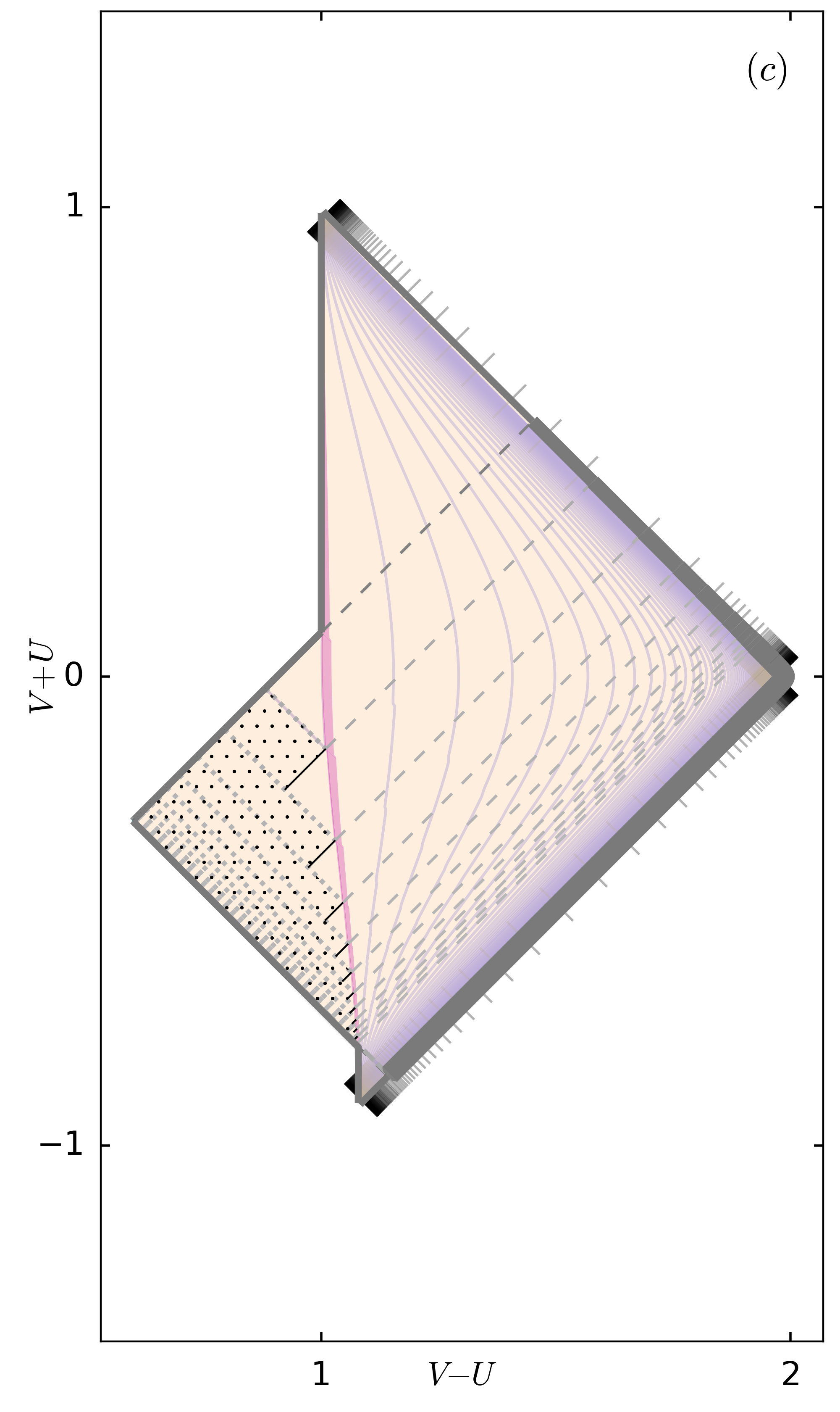}
 \end{tabular}
 \renewcommand{\plotwidth}{2.14in}
 \begin{tabular}{cc}
      \includegraphics[width=\plotwidth, trim={0 2.15in 0 0}, clip]{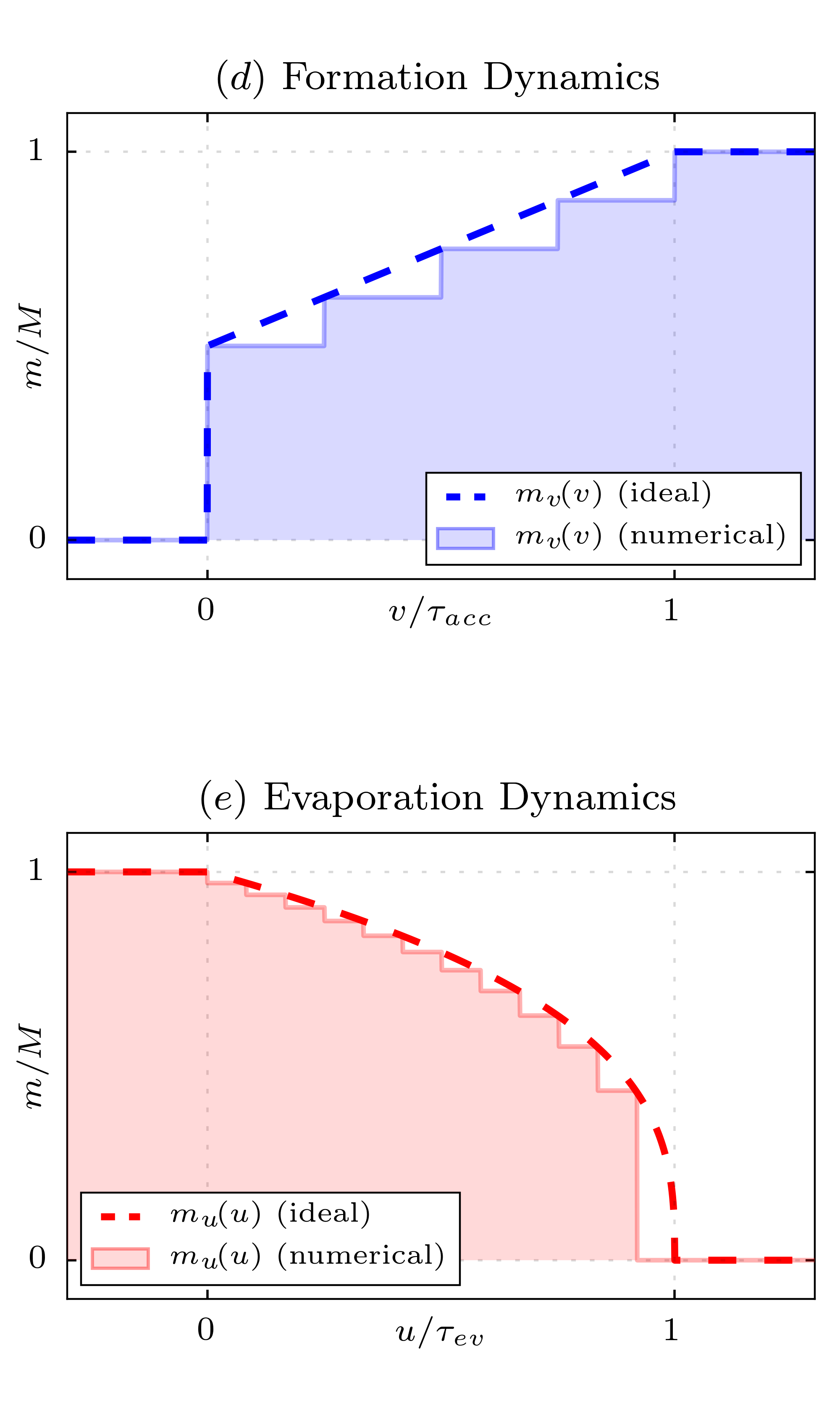}
      &
      \includegraphics[width=\plotwidth, trim={0 .15bp 0 2in}, clip]{005_mass.png}
 \end{tabular}
     \caption{
     (Color online).
     Penrose diagrams for a nonsingular Hayward black hole which forms gradually then evaporates by slowly emitting thermal radiation with a standard time dependence~(see~(\ref{eqn:mvv}-\ref{eqn:muu})). The visualization scheme and associated legend are the same as in Fig.~\ref{fig:simple-models}. 
     As discussed in Section~IV, these diagrams accurately capture both global and local (interior) structure, since they are constructed by directly finding a compact global double-null coordinate system for the spacetime.
     Although the three diagrams depicted in (a,b,c) appear vastly different, they all are derived from exactly the same \mbox{spacetime --- they} have strictly the same causal structure, and are related by conformal transformations in the $UV$ plane. (These ``conformal" transformations are induced by coordinate transformations, so no geometric information is lost, see Section~IV.) While the conformal freedom in Penrose diagrams is well known, it is not widely recognized how drastically different these transformations can make the spacetime appear, or that no single diagram will clearly depict the various widely different timescales of BH evolution (formation, evaporation, Planck scale) simultaneously. Inevitably, in any diagram, some features will be squished beyond recognition; hence the need to compare multiple diagrams to gain a full understanding of the spacetime.
     Here, different transformations are used to highlight features (a)~during accretion, (b)~during evaporation, and (c)~near the end of evaporation. 
     In comparing the causal structure between them, note that some lines appearing null are only nearly null, and that some features are hidden by being very squished. For example in (a), the entire evaporation process (which includes important timelike features visible in (b,c)) is squished into a tiny, seemingly null, line. Similarly, parts of the high density region (orange) in (b), and all of the high density region in (c), are squished and hidden behind the future part of the nearly null segment of $r=0$.
     The important qualitative features of these diagrams are summarized in Section~VIII.
     Parameters were chosen to be as realistic as numerically allowed, providing a strong hierarchy of the formation, evaporation, and interior length/time scales (an overall scale factor is irrelevant).
     Since the BH length scales are so small compared to the evaporation rate, lines $r=const$ at $2M$ (magenta) and smaller length scales are very close together and each appear as a single bundle; an additional set of $r=const$ lines has been added with spacing $dr=\tevap / 20$ (faint purple) to display larger scales.
     Although some lines of constant radius may seem discontinuous, they are in fact just strongly kinked; their continuity can be confirmed by gradually adjusting the parameters from less extreme values. For example, faint purple lines approaching the evaporating horizon in (a) do not disappear, but closely hug the horizon in a bundle until reappearing in the far future region.
     The opacity of the tick mark near the moment of evaporation in (a) shows that many tick marks have piled up there; this is the phase pileup usually associated with Hawking radiation in particle creation calculations. The same phase pileup can be observed in (b,c); in all cases one expects interesting results wherever the $dv$ ticks are very mismatched with traced-back $du$ ticks.
     Panels (d,e) confirm that the numerically generated dynamics closely approximate the desired behavior. 
     }
     \label{fig:nonsingular-realistic}
 \end{figure*}

 The desired time evolution is specified by a pair of mass functions $m_{u}(u)$ and $m_{v}(v)$ describing the mass as a function time measured by distant observers at future and past null infinity, respectively. The mass functions are mutually independent outside the requirement
 \begin{equation}
     m_u(-\infty)=m_v(\infty)=M,
 \end{equation}
 where $M$ represents the total maximum mass of the BH.
 
 The function $m_v(v)$ defines dynamics for the process of BH formation and accretion. The correct form is determined by astrophysical processes, and is of little interest to us here. As a rough estimate, we assume the BH accretes half its total mass in an initial burst at $v=0$, then accretes the remainder linearly until a time $v=\tform$ when it is fully formed:
 \begin{equation}
 \label{eqn:mvv}
     m_v(v) = 
     \left\lbrace
     \begin{array}{lc}
        0,  & v<0, \\
        M \, (\frac{1}{2} + \frac{1}{2} \frac{v}{\tform}),  & 0<v<\tform, \\
        M,  & \tform<v.
     \end{array}
     \right.
 \end{equation}
 Inspection of the diagrams reveals that during accretion, the outer horizon where $f(r)=0$ is spacelike.
 
 The function $m_u(u)$, meanwhile, defines the dynamics for BH evaporation. It is chosen to respect the thermal evaporation rate
 \begin{equation}
     \frac{dm}{du} \propto - m^{-2}
 \end{equation}
 arising from blackbody radiation calculations in a Schwarzschild spacetime~\cite{Page:1976df}. In particular, we choose
 \begin{equation}
 \label{eqn:muu}
     m_u(u) = 
     \left\lbrace
     \begin{array}{lc}
        M,  & u<0, \\
        M \left(1 - \frac{u}{\tevap} \right)^{1/3},  & 0<u<\tevap, \\
        0,  & \tevap<u.
     \end{array}
     \right.
 \end{equation}
 The location of $u=0$ is set by the requirement that Hawking pairs nucleate just outside the horizon; a distant observer sees evaporation begin at the same moment they see the infalling shells fall through their own horizon. Once evaporation begins, the horizon is timelike.
 
 While more realistic than the single-shell case, this is still only a toy model, meant to capture the commonly considered, highly idealized, case where an initially large spherical BH evaporates entirely by emitting blackbody radiation. The true spacetime of an evaporating BH is expected to differ in many ways, including corrections to the time dependence due to temperature-dependent emission effects and due to deviations from the Schwarzschild metric, as well as more intractable differences (like the necessity of including charge and rotation). Moreover, it seems likely that once an evaporating BH approaches the Planck scale its dynamics may be greatly modified. But in the absence of a widely accepted model for the BH end state, continuing Schwarzschild blackbody evaporation until the BH's disappearance seems like a conservative option. In any case, regardless of these details, this model should help attain an accurate qualitative picture of any BH emitting roughly thermal Hawking radiation during part of its lifespan.

 Choosing a sequence of shells approximating the ideal accretion dynamics is trivial, but approximating the evaporation dynamics is slightly more involved. We consider the shell approximation successful if both $m_u(u)$ and $m_v(v)$ are reasonably well approximated, with all evaporation shells nucleating at a fixed radial distance $\levap$ outside the apparent (outer) horizon $r_+$ of the region to its past (a basic assumption of our shell model). During evaporation, each shell nucleation point is defined by its coordinates $(u,v,r)$ and $(u',v',r')$ in the past and future regions containing it. Any two of $(u,v,r)$ determine the third, and the radii $r=r'$ measured on each side must be equal. Regions between evaporation shells are characterized by their mass $m$ and duration $(du,dv)$ measured by observers at future and past null infinity. A suitable sequence of shell parameters is specified as follows. As a first estimate, we sample values at equal intervals of $u \in (0,\tevap)$ from the continuous dynamics
 \begin{equation*}
     \begin{array}{rcl}
        m(u) &=& M \big(1-u/\tevap\big)^{1/3} ,\\[2pt]
        r(u) &=& 2 \, m(u) + \levap ,\\[2pt]
        v(u) &=& u + 2 \, F_S\big(r(u), m(u)\big),
     \end{array}
 \end{equation*}
 where $F_S(r,m) = r + 2m \ln \big|\frac{r}{2m}-1\big|$ is a tortoise function~\cite{Schindler:2018wbx} for a mass $m$ Schwarzschild metric, yielding a sequence $(m_i, u_i, v_i)$ of mass parameters and shell coordinates. The parameters thus obtained would be exact for a continuously evolving Schwarzschild spacetime, but are not quite consistent with our discretized (and possibly non-Schwarzchild) model; this is corrected by adjusting the $u_i$ values to ensure that nucleation points lie exactly $\levap$ outside the outer apparent horizon. This process determines the discrete dynamics up to an overall time translation. One might expect that the small adjustment of $u_i$ always leads to a valid approximation of the ideal dynamics, but in practice the method can break down if the BH mass and lifetime are not mutually consistent.%
 \footnote{Page has shown based on quantum and thermodynamic analysis that for an evaporating Schwarschild BH one expects
 $$\left(\frac{\tevap}{\tpl}\right) = A \, \left(\frac{M}{\mpl}\right)^{\! 3} ,$$
 where $A \sim 1000$ is a unitless constant determined by physical considerations~\cite{Page:1976df}. (The precise value of $A$ depends on various factors since the proportionality constant in the evaporation rate is, in less idealized cases, temperature dependent.) Our model seems to work well unless the nominal evaporation time is much shorter than dictated by this relation, with the Planck scale set by $\levap$ (in geometric units $\levap = \lpl = \tpl = \mpl$). Since our model is purely geometric, the reason this relation must be enforced is not trivial.
 } %
 Whether the approximation was successful can be checked empirically; Figs.~\ref{fig:nonsingular-realistic}d--\ref{fig:nonsingular-realistic}e demonstrate that the desired continuous dynamics was correctly attained for the spacetime of Fig.~\ref{fig:nonsingular-realistic}.

\begin{figure}[t]
     \includegraphics[scale=1]{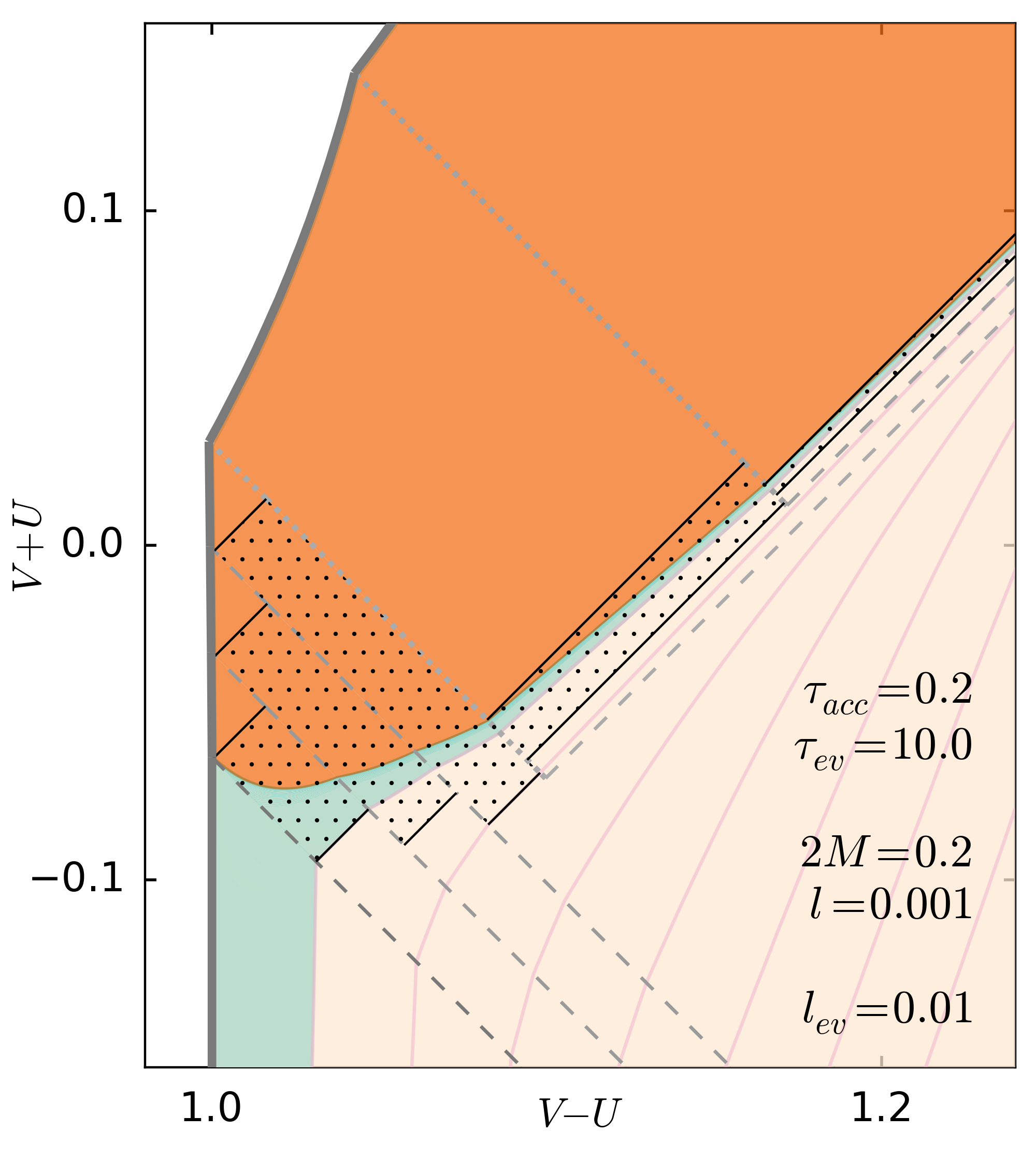}
     \caption{%
     (Color online). Detail view of horizon and core dynamics during formation and the early stages of evaporation, for a spacetime similar to Fig. \ref{fig:nonsingular-realistic} but with somewhat less realistic parameters (in the sense that the mass and lifetime are not mutually consistent). Although the parameters are less realistic, the features depicted are qualitatively accurate. Details like these are not visible in more quantitatively accurate diagrams, where the extreme hierarchy of length and time scales prevents accretion and evaporation features from being depicted simultaneously. Compare to the region where accretion shells meet the core in Fig. \ref{fig:nonsingular-realistic}a. Note that the apparently teal region is in fact a bundle of closely spaced lines of constant radius just outside the core.
     }
     \label{fig:zoom}
 \end{figure}

 Three different diagrams are needed to visualize different aspects of the spacetime in Fig.~\ref{fig:nonsingular-realistic}, due to the presence of vastly different timescales in the problem. Each panel corresponds to a different (conformally related) coordinate transformation in the $UV$ plane, depicting the process as ``viewed" from one of three different times. (The exact form of the coordinate transformations is determined by which region is used as the ``seed" for a chain of shell matching transformations.) Together, they paint an intuitive picture. Early observers see the BH accrete mass, forming a BH with a long-lived nearly null horizon, behind which lies the dense core (space between the horizon and core exists but is squished away in this diagram). Intermediate-time observers looking far into the past see the outer surface of accretion shells shrouded by a horizon, with sparse Hawking radiation to their past and a final evaporation blast in their future. And late time observers simply see Hawking radiation emitted from a timelike horizon. The detail view of Fig.~\ref{fig:zoom} emphasizes observers near the horizon at early times, and shows clearly the horizon transitioning from spacelike to timelike as accretion gives way to evaporation. 
 
 In the following section, we will see that a similar model can be extended to include cosmological models with nonzero background curvature.

\renewcommand{\plotwidth}{1.6in}
 \begin{figure}[t]
 \begin{tabular}{cc}
      \includegraphics[width=\plotwidth]{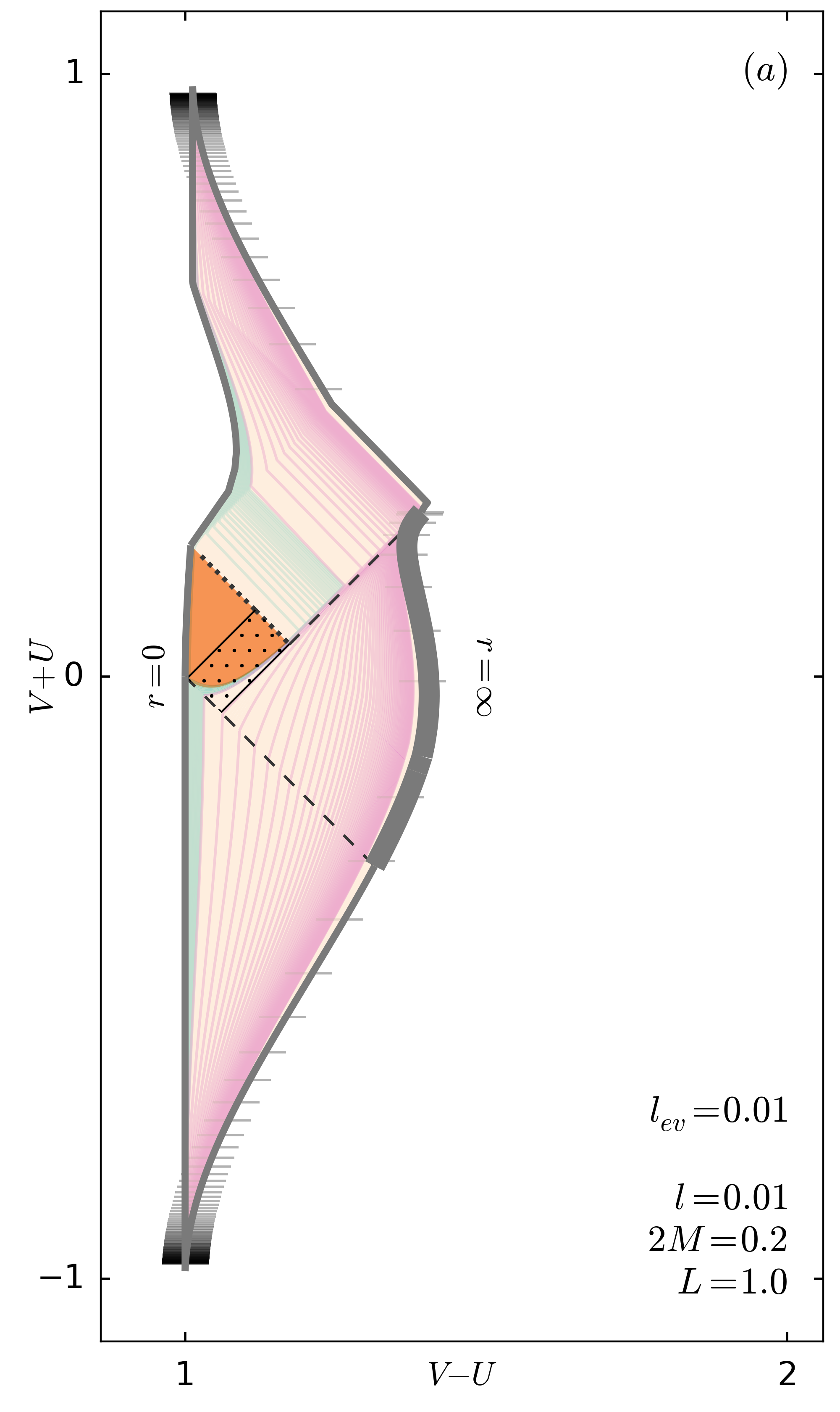}
      &
      \includegraphics[width=\plotwidth]{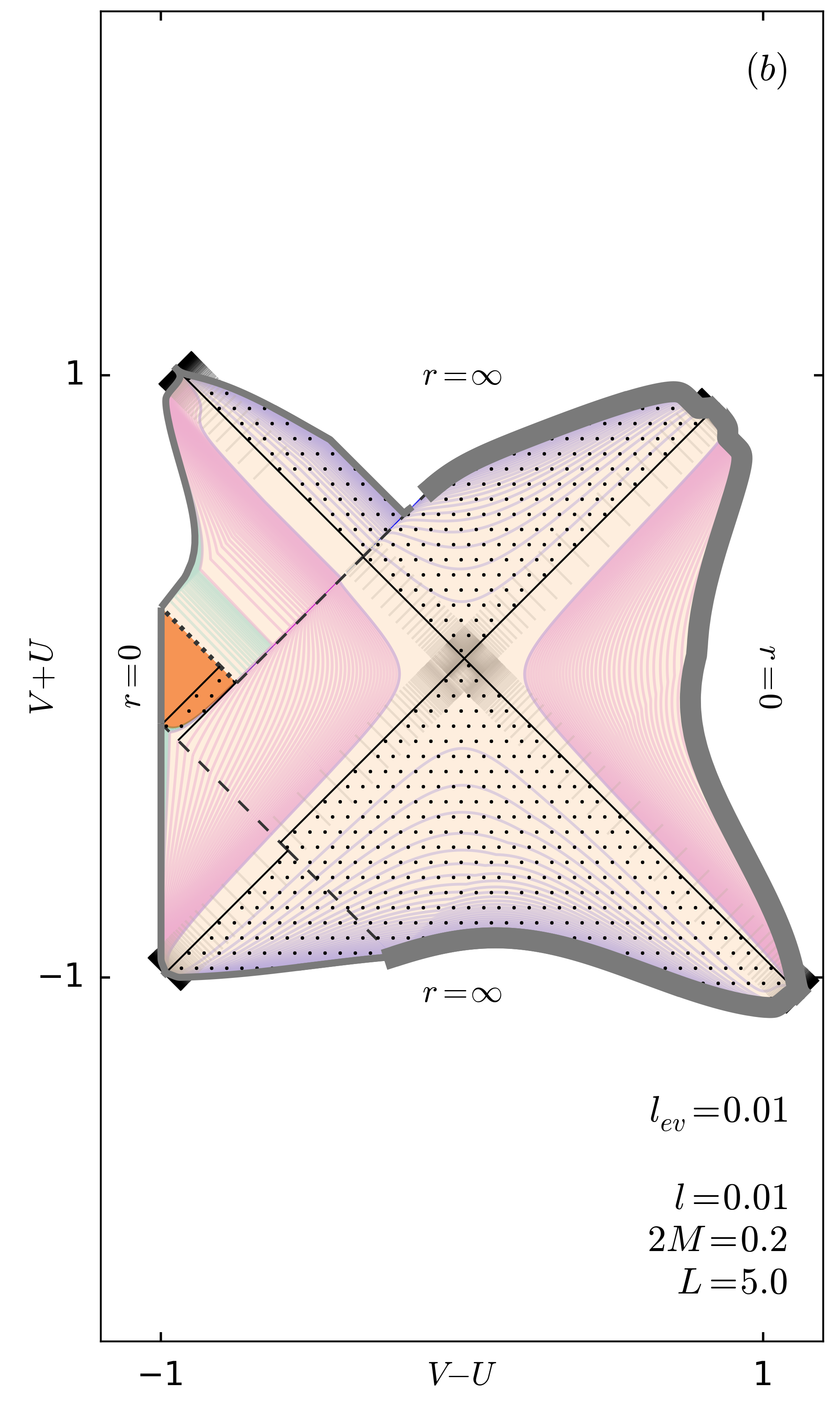}
  \end{tabular}
     \caption{(Color online). Penrose diagrams for single-burst shell models with (a) negative and (b) positive background curvature, with metrics (\ref{eqn:hay-AdS-metric}) corresponding to the Hay-AdS and Hay-dS BH spacetimes. Parameters were chosen for maximum visibility of qualitative features. Density color scale here depicts $\rho - \rbg$, with constant background density $\rbg=\pm 3/(8\pi L^2)$ (positive for dS). In~(b), an additional set of lines $r=const$ (faint purple) have been added with spacing $dr=L/2$ to depict larger length scales. Note that, in addition to BH trapping, the dS case includes cosmological trapped regions having nothing to do with the BH. Tick marks no longer correspond to proper time, but still represent equal increments of $du$ and $dv$. Notably, inspecting the tick marks shows that the phase pileup typically associated with Hawking radiation is still present in the $u$ and $v$ coordinates, even without the assumption of asymptotic flatness.}
     \label{fig:background-curvature}
 \end{figure}

\section{VII. \ \ Diagrams including\\ background curvature}

While many discussions of BH evaporation assume, for simplicity, an asymptotically flat exterior, it is also interesting to consider BHs evaporating in alternative cosmologies. We therefore provide, in this section, Penrose diagrams for nonsingular BHs evaporating within asymptotically de Sitter (dS) and Anti de Sitter (AdS) spacetimes. Let us suppose, for both dS and AdS cases, that the background curvature is characterized by a length scale~$L$ such that $l \ll 2m \ll L$ ($l$ and $m$ being parameters of a Hayward metric, see above). That is, the black hole is small compared to background curvature length~scales.

A nonsingular BH in a background of constant curvature may be described by (\ref{eqn:sss-metric-0}) with
\begin{equation}
 \label{eqn:hay-AdS-metric}
     f(r) = 1 - \frac{2m r^2}{2m l^2+r^3} \pm \frac{r^2}{L^2},
\end{equation}
which we refer to as the Hay-AdS~($+$) and Hay-dS~($-$) metrics.
By fixing $l$ and $L$ while varying $m$ across shells, we can construct single-burst forming and evaporating black hole models analogous to the flat space version of Fig.~\ref{fig:simple-models}b. The results are depicted in Fig. \ref{fig:background-curvature}.%
\footnote{%
Maximally extending BH metrics in a dS background generically leads to an infinite chain of BHs and asymptotic infinities. To avoid this issue we modify the Hay-dS spacetime to transition to pure dS space far away from the BH. The transition occurs inside the cosmological horizon on the side of dS with no BH, and has no important effects.
} %

Having constructed single-burst models, the question arises of how to properly add continuous time dependence. Assuming $2m \ll L$, the spacetime is nearly asymptotically flat on length scales much smaller than~$L$, and there is a class of observers, similar to distant Schwarzschild observers, for whom $2m \ll r_{\rm obs} \ll L$ and $f(r) \approx 1$. It seems reasonable to assume that these nearly-asymptotically-flat observers should measure the usual thermal dependence, in which case the earlier prescription in terms of $m_u(u)$ and $m_v(v)$ carries over unmodified. For a cosmologically-sized BH, however, how the detailed structure would look is less clear.

\section{VIII. \ \ What is a black hole?}

The causal structure encoded in these diagrams invites us to revisit a subtle question: \textit{What is a black hole?} A~number of authors, most famously Hawking~\cite{Hawking:2014tga}, have already \evilpad{0.1pt}{argued}{0.5pt} that the traditional global \mbox{definitions} (e.g. in terms of an event horizon) are not useful when evaporation is taken into account --- and we will make a more detailed case for this proposition below. While the natural context for defining black holes is quantum gravity, to the extent that BHs have a classical spacetime description, a general relativistic definition should be possible. To motivate an improved definition, let us review some of the features of the spacetime of Fig.~\ref{fig:nonsingular-realistic}:
\begin{enumerate}[(i)]
    \item The spacetime has no event horizon, and is globally hyperbolic.
    \item A distant external observer sees collapsing matter fall through its own apparent horizon in a finite amount of (the observer's) proper time. At the same moment the observer sees this crossing occur and the apparent horizon form, they begin to receive Hawking radiation (compare Figs.~\ref{fig:simple-models}b,~\ref{fig:nonsingular-realistic}a,~\ref{fig:zoom}). In general, distant observers see Hawking radiation if and only if they are looking back at the apparent horizon.
    \item Hawking radiation is emitted from a timelike surface (Fig.~\ref{fig:nonsingular-realistic}c). The emission surface is just barely outside the apparent horizon, which itself is also a timelike surface in the continuous limit (Fig.~\ref{fig:nonsingular-realistic}c).
    \item The part of the diagram intuitively considered ``part of the BH" consists of the trapped region and core. The trapped region is ``trapped" in the sense that all future-directed curves are radially ingoing. The core is an ultradense quantum gravity plasma.
    \item During evaporation, the outer boundary of the trapped region is a timelike apparent horizon (Fig.~\ref{fig:nonsingular-realistic}c). During formation, a spacelike portion of the apparent horizon occurs inside the infalling matter (Fig.~\ref{fig:zoom}). External observers previous to the emission of any Hawking radiation see the apparent horizon as a nearly null surface in the future (Fig.~\ref{fig:nonsingular-realistic}a); in order to receive Hawking radiation, such observers must go ``around the corner" to where the horizon appears timelike.
    \item Some trapped observers, who have just barely fallen in, can escape the BH without intercepting the core by accelerating out of the trapped region during evaporation. Others, who have fallen too far in already, are doomed to destruction in the core. This ``region of no escape" is quantified by the past domain of dependence of the core (the location of which can be inferred from Fig.~\ref{fig:nonsingular-realistic}b). The core surface is spacelike, while the boundary of the doomed region is null.
    \item The inner boundary of the trapped region (the inner horizon) lies entirely within the core; it is timelike during accretion and spacelike during evaporation (Fig.~\ref{fig:zoom}). In the final moments of evaporation, the inner horizon, core surface, and outer horizon all come together.
    \item The core maintains a constant Planck scale density. As the BH mass changes over time, the core, which for large BHs is significantly larger than Planck scale in radius, adjusts in size to accommodate the total mass.
    \item The proper time experienced at $r=0$ within the core (to the extent that it corresponds to the classical value) is of the same order of magnitude as the BH lifetime experienced by distant observers. However, this could potentially be modified by altering the Hayward metric to include a redshift factor (see discussion surrounding (\ref{eqn:redshift})).
    \item Time evolution starting from initial data in the distant past and proceeding through the process of BH formation, accretion, and evaporation down to the Planck scale (including the emission of the majority of the Hawking radiation) can be described by a continuous family of Cauchy surfaces entirely to the past of the (quantum gravity) core (this is seen most clearly in Fig.~\ref{fig:nonsingular-realistic}c), on which semiclassical physics should apply. Evolution beyond the final moment of evaporation, when the BH disappears entirely, involves evolving Cauchy surfaces through the core, and requires a quantum gravitational description.
\end{enumerate}

Inspired by the above observations, we propose a definition which depends on only local quantities and is consistent with all common black hole models: \textit{A black hole is a future-trapped region surrounding and feeding into an ultra-dense core}.%
\footnote{%
The phrase ``feeding into" signifies that the inner future boundary of the trapped region lies on or within the core, and the phrase ``surrounding" implies that the core lies within future light-sheets \cite{Bousso:1999xy} of the relevant closed trapped surfaces. Recall also that the trapped region is ``trapped" in the sense of trapped surfaces, not of ``no escape."
} %
Both the core and trapped region should be considered a part of the BH; one might propose to call this a \textit{core and shroud} definition. Insofar as a singularity acts as a placeholder for a dense point mass, this definition includes both singular and nonsingular models.

This definition departs from tradition, by focusing on the trapped region and apparent horizon, rather than the region of ``no escape to infinity" and corresponding event horizon. The traditional ``event horizon" definition has several disadvantages. Most importantly, the location of an event horizon (which is typically defined as the boundary of the past of future infinity) is determined by the entire future history of the BH, and cannot be determined by any local data. Thus, even when it is clear that a ``BH-like object" locally exists, the presence or absence of a BH by the traditional definition depends strongly on the BH end state (for example whether the BH later accretes additional mass, and whether the final dynamics are dominated by evaporation, remnant outcomes~\cite{Rovelli:2014cta}, mass inflation instability~\cite{Poisson:1989zz}, or a bounce~\cite{Haggard:2014rza}). In contrast, the location of a core and trapped region can be determined by local data in a finite time.

Moreover, even without invoking nonsingular models, it's not clear that rotating or charged BHs form an event horizon at all when evaporation is taken into account. For example, the naive translation of Fig.~\ref{fig:traditional-diagram} to a Reissner-Nordstrom metric leads to a naked singularity with no event horizon (such a diagram looks similar to Fig.~\ref{fig:simple-schematic}b, see,~e.g.,~\cite{Kaminaga:1988pg,Parikh:1998ux}). One possible rebuttal, that BHs should discharge and spin down before evaporating, is not very convincing, since spacelike-ness of the Schwarzschild singularity is unstable to even continuously small perturbations of the charge and rotation parameters. If nonsingular models are adopted, then charged and (presumably, see~\cite{Burinskii:2001bq}) rotating BHs have the same basic causal structure as the Hayward metric, and, as in Fig.~\ref{fig:nonsingular-realistic}, would exhibit no event horizon in our simple model. General nonsingular solutions should be expected to have a causal structure which is stable under perturbations of the rotation and charge, and to alleviate the pathologies of the interior Kerr-Newman metrics.

In the new definition, some conceivable BHs (which may or may not actually exist in nature), like the Schwarzschild BH of Fig.~\ref{fig:simple-models}a, have an event horizon, while others do not. In the simplest case with an event horizon, where a Schwarzschild BH forms quickly then slowly evaporates without disturbance, the null event horizon is just barely inside the timelike apparent horizon (this fact can be inferred from Figs.~\ref{fig:simple-models}--\ref{fig:nonsingular-realistic}, and checked numerically); the horizons nearly coincide. In other cases, for example if the BH forms, evaporates for some amount of time, then later accretes more mass, the event horizon and apparent horizon can be widely separated.

In any case, at least in our classical model, energy conservation by the DTR relation forces Hawking radiation to be emitted from just outside locations where $f(r)=0$ (in the Schwarzschild metric, $r=2m$), so that Hawking radiation emission is directly tied to the boundary of the trapped spheres region. The boundary of the trapped spheres region is also the apparent horizon (see appendix), so the radiation can be thought of as emanating from the apparent horizon.

In any classical model like our shell model, therefore, the Hawking radiation emission points must not be tied to the event horizon if they are to conserve energy. For instance, in the second case above, where a Schwarzschild BH forms, evaporates for some amount of time, then later accretes more mass, \textit{the early Hawking radiation would be emitted from far inside the event horizon}. Such radiation would not make it to infinity, but could still be observable to distant observers for an arbitrarily long time before the second accretion event takes place. This observation provides further evidence that the apparent horizon is a more physical candidate than the event horizon to describe the BH boundary, motivating the new definition.

As discussed in the appendix, our trapped spheres region is the spherically symmetric special case of the more general ``trapping nucleus," which is an essential subset of the trapped region. It is conjectured that in general there is a distinguished trapping nucleus whose boundary is a uniquely defined apparent horizon~\cite{Bengtsson:2010tj}. It has been argued from a trapped surfaces perspective that such a trapping nucleus provides the most reasonable definition for a BH and its boundary~\cite{Bengtsson:2010tj}. The above observation about energy conservation in our shell model acts as an independent check supporting this hypothesis, making it likely that in similar but not-spherically-symmetric semiclassical models, Hawking radiation emission would need to come from just outside a trapping nucleus to conserve energy. Also, this nucleus is likely easier for an observer to identify than the full trapped region, since the nucleus consists of relatively trivial trapped surfaces and its boundary is marginally trapped.

While the nonsingular BH of Fig.~\ref{fig:nonsingular-realistic} does not have an event horizon or region of no escape, the past domain of dependence of the core, which might be called the ``doom~region," plays a similar role. Any observer crossing into the doom region will inescapably be crushed into the ultra dense quantum gravity core before being emitted in the Hawking radiation. Notice that the boundary of the doom region in Fig.~\ref{fig:simple-models}b almost exactly coincides with the event horizon in Fig.~\ref{fig:simple-models}a, a phenomenon that holds generally when a Schwarzschild metric is replaced by Hayward. In this context, the singularity theorems show that doom regions are generic: matter which collapses through its own apparent horizon will inevitably continue collapsing until quantum gravitational effects kick in, forming a core and associated doom~region.

By focusing on the properties of a BH as an actual compact object that can form and exist at a finite time, the new definition allows a broader and more useful class of objects to be called BHs, like evaporating charged, rotating, and nonsingular metrics, BH-like objects which later bounce or form a remnant, and collapsed stars whose exact metric and eventual end state is unknown. In turn, this allows the study of when certain pathologies of BH metrics, like singularities, event horizons, and Cauchy horizons, do and do not arise. Defining BHs (which we fundamentally know to exist only as compact astrophysical objects) by insisting on their worst pathologies is one reason that questions of unitarity and information preservation during evaporation have remained unresolved.

\section{IX. \ \ Towards a self-consistent evaporation model}

Significant attention has been given recently to the question of self-consistent BH evaporation models (related examples include \cite{Kawai:2013mda, Kawai:2017txu, Chen:2017pkl, Frolov:2016gwl, Frolov:2017rjz}). Typically the idea is to postulate an evaporating BH spacetime, treat it as a fixed background for a quantum field theory, and show that a renormalized stress tensor for the field theory matches the background curvature --- or at least find some evidence that the field theory and background are compatible. It is our hope that the present diagrams, especially Fig. \ref{fig:nonsingular-realistic}, can help inform this effort.

In particular, for example, the model presented here suggests an improvement to the recent interesting calculation by Frolov and Zelnikov, who have studied the quantum radiation from an evaporating modified Hayward metric \cite{Frolov:2016gwl,Frolov:2017rjz}. They found, in addition to the Hawking radiation, an unwanted burst of radiation emanating from the inner horizon. Moreover, the burst was found to be at least partially mitigated by certain changes to the metric. Our model suggests even further changes to the metric. Specifically, time-dependence in this study was included only by virtue of a function $m(v)$ depending on the time parameter at past null infinity, which was assumed to have the usual thermal-evaporation time dependence. This has two drawbacks. First, it is observers at \textit{future} null infinity, \textit{receiving} the Hawking radiation, who should measure the correct thermal time dependence~---~not those in the past. That is, it is $m(u)$, not $m(v)$, that should be thermal during evaporation. And second, using only $m(v)$ implicitly assumes that evaporation occurs by absorption of negative mass shells \textit{incident from infinity}, when they should rather be incident from near the horizon. A more physical metric would include a more complicated time dependence, depending on both $u$ and $v$ simultaneously. The proposed change is drastic enough to hope that, coupled with an expeditious choice of redshift function, it could help the model approach self-consistency.

Related to the study of self-consistent models, and to the unwanted energy outburst discussed above, is the phenomenon of mass inflation, in which large amounts of gravitational energy are converted into mass by the collision of null shell perturbations in a BH \cite{Poisson:1989zz}.

\begin{figure}[t]
     \includegraphics[scale=.63]{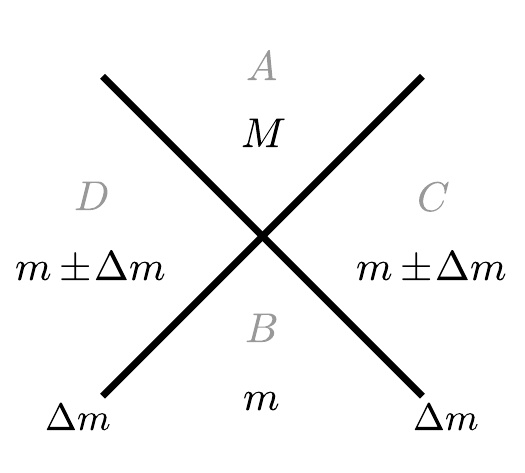}
     \caption{Two spherical null shells colliding in a spherically symmetric spacetime. Gray lettering indicates the standard labelling of each region in DTR calculations (see appendix). Black lettering describes an example: Two shells, each of mass~$\Delta m$, collide, in an initial Schwarzschild spacetime of mass~$m_B=m$. Assuming $A,B,C,D$ are all Schwarzschild spacetimes, the masses $m_C$ and $m_D$ are fixed by shell junction conditions, with signs determined by whether each shell is radially ingoing ($+$) or outgoing ($-$) towards the future. The final mass $m_A = M$ is determined by the DTR relation. Assuming all shells (both initial and final states) have a positive mass, mass inflation can occur only if both shells are radially ingoing --- for instance, inside the trapped region of a BH. Otherwise, restricting to positive-mass shells ensures that $|M-m|\leq \Delta m$.
     }
     \label{fig:shell-collision}
 \end{figure}

The cause of mass inflation can be understood as follows. Consider the shell collision depicted in Fig.~\ref{fig:shell-collision}. Two spherical null shells, each of mass $\Delta m$, collide, with initial and final Schwarzschild masses $m$ and $M$ respectively. If the collision occurs \textit{outside} the BH horizon, where future-directed shells come in a radially-ingoing/radially-outgoing pair, then the requirement that all shells have positive mass ensures that $|M-m|<\Delta m$. This makes sense since an ingoing (outgoing) positive-mass shell will increase (decrease) the mass of a Schwarzschild region to its future, so that the shell masses roughly cancel out in the final state. But if the collision occurs \textit{inside} the black hole, both incident shells are radially ingoing: they both contribute positively to the final mass. Working out the associated constraints shows that in this case, even when all shell masses are positive, the final mass $M$ can be arbitrarily large. When both incident shells are ingoing and the collision occurs near $f(r)=0$ (i.e. near a horizon), application of the DTR relation (see appendix) shows that even very small incident masses will lead to an unbounded increase in $M$: this allows the mass inflation.

There is an open question of whether mass inflation at the inner horizon creates an instability which significantly alters the dynamics or interior metric~\cite{Hamilton:2008zz}. Our model suggests a few more remarks on this topic.

First, the mass inflation instability is often associated with a part of the inner horizon, the ``outgoing" inner horizon~\cite{Hamilton:2008zz}, which does not actually exist in evaporating BH spacetimes like ours or Hayward's. This outgoing horizon, which acts as the Cauchy horizon in simple collapse models, is the part of the inner horizon approached by right-directed (in the standard diagram, see Fig.~1~of~\cite{Poisson:1990eh}) null rays in an eternal charged or nonsingular BH metric. (These rays are technically ingoing despite the nomenclature.) This part of the horizon exists in the eternal metric, but is cut off by an incoming negative-mass shell in our Fig.~\ref{fig:simple-schematic}b, and cut off by the smooth evaporation process in Hayward's model. The evaporating models also have no Cauchy horizon for the same reason. While it is not impossible for a right-going shell in our model to approach the inner horizon, such shells typically emanate directly from an earlier part of the same inner horizon, and do not intersect the ``outgoing" part of the horizon as usually assumed. 

This is not an argument that mass inflation does not occur. But it does lead to a delicate situation. It is common to study mass inflation within the causal structure of an eternal BH spacetime where the outgoing inner horizon exists (for a clear example see~\cite{Carballo-Rubio:2018pmi}). But if an evaporating spacetime is assumed, some of the assumptions underlying these calculations may break down. On the other hand, it has previously been argued both that mass inflation does not depend on the long term future~\cite{Hamilton:2008zz}, and that evaporation should fail to prevent mass inflation once corrections to the Schwarzschild evaporation rate are taken into account~\cite{Carballo-Rubio:2018pmi}, possibly circumventing the issue we are raising. However since both still make the assumption described, the resolution may or may not be definitive.

Second, if curvature is assumed to be regulated by a Planck scale density cutoff, then in nonsingular models with sufficiently low charge and rotation (like the Hayward metric), the entirety of the inner horizon is hidden far within the quantum-gravity-dominated core. Any calculations based on semiclassical physics near the inner horizon must therefore be called into question in these cases. Further, if the result of mass inflation is meant to be the development of an extreme density at the inner horizon~\cite{Poisson:1990eh}, it is not clear that any such higher density could be reached. This disrupts the interpretation of mass inflation in static nonsingular cases (as in, e.g.~\cite{Carballo-Rubio:2018pmi}). On the other hand, it is possible that rotating and charged nonsingular BHs have an inner horizon extending outside the core~\cite{Burinskii:2001bq}, so the issue still must be dealt with unless there is a sufficient mechanism for discharge/spin-down.

\section{X. \ \ Possible implications for the physics of black hole evaporation}

The questions we feel it natural to ask about BH evaporation can change drastically depending on what spacetime diagrams are used. In the traditional diagram, for example, it's natural to ask: \textit{Is there a unitary transformation connecting the quantum states at future and past null infinity?} But in the nonsingular evaporation diagrams presented above, which have been constructed to be globally hyperbolic, it's almost obvious that such a transformation ought to exist between any two Cauchy surfaces not intersecting the ultra-dense quantum gravity region (where it doesn't seem safe to assume known physics applies). These new diagrams, on the other hand, raise a different fundamental question: \textit{What is the nature of the negative-mass shells?}

A global, phenomenological, and ultimately unfulfilling answer to the latter question already exists. As shown most clearly by Davies, Fulling, and Unruh~(DFU), quantum field theories yield an ingoing negative energy flux in a renormalized stress tensor near the horizon when a BH is present \cite{Davies:1976ei}. Clearly, the presence of this ingoing negative flux (and of our ingoing negative-mass shells) is a phenomenon of quantum field theory in curved spacetime. But the calculation is global. Can this phenomenon be understood as a local process? How should the negative flux be interpreted physically?

Suppose we answer these questions with a bold but simple interpretation: that the negative-mass shells represent normal matter propagating out from the core on non-future-directed (i.e.\ past-directed or spacelike) trajectories.%
\footnote{%
A classical connection between negative mass and non-future-directed propagation is fostered by calculating the energy flux vector $F^\mu = - T^{\mu\nu}  \,t_\nu$ relative to a timelike observer for a uniform dust of density $\rho$ with velocity $u^\mu$ and stress tensor $T^{\mu\nu}=\rho\,u^\mu u^\nu$. Both $\rho<0$ and $u^2 > 0$ lead to non-future-directed energy flux.
} %

With a little justification, this seems like a reasonable claim. It is well known that quantum field theories~(QFTs) violate, by small amounts, the classical energy conditions, including energy conditions intended to prevent the non-causal transfer of energy~\cite{Martin-Moruno:2017exc}. The fact that these theories violate the energy conditions is closely tied to the faster-than-causal spreading of relativistic wavepackets in the one-particle sector of QFT~\cite{Rosenstein:1987a,Redmount:1990mi}. In this sense, energy condition violations are related to a small amplitude for QFT particles to propagate, or in a probability interpretation ``tunnel," faster than light. For there to be a non-negligible probability of escape from the core by this process, wavepackets must spread beyond the trapping radius, so that escaping field modes have wavelength on the order of the BH size. As the BH gets smaller, the amount of non-causal propagation needed to escape is reduced, and the process becomes more probable, so evaporation speeds up. Normally the non-causal propagation is overwhelmed by much more likely causal propagation, but a BH makes escape by causal propagation impossible.%
\footnote{It must be noted that using any language of ``particles" (or even ``quanta") is extremely dangerous in this context, due to the many well-known ambiguities surrounding QFT particles in curved spacetime~\cite{Birrell:1982ix}. The field picture must be given conceptual priority. If not taken too literally, however, particle language can sometimes provide a useful heuristic. When we use the particle terminology here, it is mainly to make connection with standard language of the literature.} %

In case of the DFU stress tensor, energy conditions are violated due to an ingoing flux of negative mass into the horizon, which, in this picture, would correspond to a non-future-directed transfer of energy out from the BH core. While it's not unusual for QFTs to violate energy conditions, what's unusual about BH evaporation, in this picture, is that non-future-directed propagation dominates the dynamics. (In Fig.~\ref{fig:nonsingular-realistic}, this refers to matter being carried out on the negative mass shells, instead of emerging through the future boundary of the core.) The fundamental question, then, is shifted to one that more directly implicates the paradoxical nature of BH evaporation: \textit{Why does matter escape from the core exclusively, or at least primarily, on non-causal trajectories?} 

The answer to this question might be more straightforward than it seems. As discussed already, even under normal circumstances QFTs violate energy conditions in ways that may allow energy to locally propagate on non-causal trajectories.
If the lightcone of some matter is blocked by a strong space-and-time-dependent potential barrier, can causal propagation be significantly depressed? If so, could the small energy-condition-violating flux add up to a significant transfer of mass outside the lightcone? And does the BH core metric (specifically, the rapidly changing metric where the extremely dense core meets the future post-evaporation vacuum, see Fig.~\ref{fig:nonsingular-realistic}) act as such a barrier when coupled to a quantum field? (\mbox{Alternately},%
\footnote{In the semiclassical model we are suggesting, $\langle T_{\mu\nu} \rangle$ corresponds to a scenario where matter exits the core on spacelike trajectories, as opposed to propagating directly through the future boundary of the core. One way to make sense of this may come from quantum gravity. Consider, heuristically, a path integral formulation for the complete quantum gravitational system, and consider some tunnelling calculation where a nonsingular BH tunnels into an evaporated state at a later time. What paths (in some quantum gravity configuration space) contribute to this process? There are some paths where matter in the core propagates timelike while the metric remains static --- these paths correspond to a purely classical evolution, but don't contribute to the tunneling process, because they only contribute to an ``eternal BH" configuration of the metric. That is, paths with timelike matter propagation in the core correspond to classical scenarios where matter sits in the core forever, not situations where matter leaves the core and the BH evaporates. Among paths where the metric looks like an evaporating BH, it's possible that paths with spacelike propagation are less suppressed than other off-shell contributions, for example less suppressed than configurations that drastically violate Einstein's equation. If this were the case, one could think of quantum gravity as creating a quantum correlation between BH evaporation in the metric, and spacelike propagation in the matter fields.
} %
could this process be a phenomenon of quantum gravity?) These questions have not yet, to the authors' knowledge, been addressed; they should be.

This line of reasoning is open to an obvious criticism: The presence of a Hawking flux depends (in curved-spacetime quantum field theory calculations) only on the presence of the horizon, and therefore cannot depend on physics in the core. This is an important observation, but, although it may seem so at first glance, it is not especially damning to the interpretation. Actually, the only details of core physics relevant to this discussion are two assumptions already implicit in singular BH models: that once an apparent horizon forms, gravitational collapse to a singularity (core) is inevitable; and that future-directed propagation out of the singularity (core) is not allowed. Moreover, if BH evaporation is truly unitary, the matter emitted in Hawking radiation should be, in some sense, ``the same" matter that formed the BH in the first place. This point is almost always neglected in discussions of particle creation by a horizon. 

The key to reconciling the fact that the Hawking flux depends only on the presence of the horizon, and the fact that matter emitted in Hawking radiation should come from the core, is self-consistency between the background spacetime and the quantum field theoretic stress tensor. This is closely tied to the question of why BH horizons supposedly evaporate, while Rindler horizons (for accelerating observers in flat space) do not. 

In the Rindler vacuum, the renormalized semiclassical stress tensor $\langle T_{\mu\nu}\rangle$ bears no resemblance whatsoever to the Einstein tensor of the flat background \mbox{metric~\cite{Birrell:1982ix,Candelas:1977zza}}; it should not, therefore, be regarded as a solution of the joint matter-gravitational field equations. But in BH evaporation, the renormalized stress tensor for the in-vacuum closely resembles the Einstein tensor of our shell~model.%
\footnote{Statements like these are ambiguous when there is ambiguity in the stress tensor renormalization scheme. For the present qualitative purposes, suppose the field equations aren't satisfied unless some reasonable renormalization scheme is shown to produce a stress tensor equivalent to the metric.
} %

One can imagine an iterative process, where the background spacetime is perturbed to approach the renormalized QFT stress tensor at each step, converging on a self-consistent model of an evaporating BH spacetime coupled to the matter field. Starting such a process from a non-evaporating BH spacetime in the usual in-vacuum, one obviously expects the horizon to start evaporating during the iterations due to the ingoing negative flux. But there is no reason to think that this should be the only effect, or that an initially vacuum region of spacetime should be empty when the process converges. (Note that ``initially" here refers to iterations, not to time.) Presumably, assuming some usual in-vacuum state, the presence of the BH horizon is \textit{not only} sufficient to predict Hawking-like radiation, \textit{but it is also sufficient} to predict the existence of a core which forms and then evaporates.

How does this observation apply to Rindler horizons in Minkowski space, and cosmological horizons in de Sitter? Let's look closer at the analogy. First of all, from a non-technical standpoint, it makes no sense for either of those horizons to generically emit radiation,%
\footnote{This is separate from, but often confused with, the question of radiation experienced by an accelerating particle detector. Particle detector calculations (see~\cite{Birrell:1982ix} for review) like the famous results of Unruh~\cite{Unruh:1976db} and Gibbons and Hawking~\cite{Gibbons:1977mu} do not deal with objective radiation existing in the spacetime and quantum state, but rather with the coupling a of a local detector to a quantum field. This conceptual distinction was made very clear, for example, by Padmanabhan and Singh~\cite{Padmanabhan:1987rq}. The excitations of such a detector should be attributed to a combination of two effects: the difference between the global quantum state (perhaps a global ``vacuum") and an observer's local inertial vacuum (that is, some state with no particles in modes defined by an observer's local inertial frame); and the observer's acceleration relative to that frame. Davies and others have suggested a useful paradigm: that the detection of particles due to acceleration be thought of as a form of ``vacuum friction"~\cite{Davies:2005ei,Manjavacas:2010aa}. The relation of particle detector calculations to thermal emission by a horizon is evident only in special cases, where constant acceleration or a particular choice of positive frequency modes creates a useful analogy.
} %
since both of the spacetimes are homogeneous --- every point in de Sitter spacetime lies on a cosmological horizon. More technically, the emission or not of radiation, and the evaporation or not of the horizon, depends on the quantum state. This is obvious, but there is a key point missing from standard discussions of this issue. States where the renormalized stress tensor is highly mismatched from the background Einstein tensor \textit{are probably never realized}, since they are not likely to be solutions of the joint matter-gravity field equations (whatever that means in quantum gravity). Yes, there are states where Rindler, cosmological, or other Killing horizons emit radiation. But they are not solutions of the joint equations of motion unless an iterative process is performed to match the background metric with the stress tensor. Perhaps these states converge to BH solutions, perhaps to something else, or perhaps they don't converge at all. Either way, in the special case of Rindler, cosmological, etc.\ horizons, the first iteration step is likely to drastically change the qualitative picture and call the horizon interpretation into question. In contrast, iteration towards self-consistency in the BH case reinforces the qualitative picture of a BH evaporating while emitting Hawking radiation. The BH case, unlike the other cases, appears to admit a \textit{self-consistent} picture of an evaporating horizon.

It seems likely that, starting from an evaporating nonsingular spacetime like Fig.~\ref{fig:nonsingular-realistic}, the iterative process described above has at least a pretty good chance to converge to something reasonable, given the qualitative similarity between the DFU stress tensor and the stress tensor for our shell model. It's not clear that the same can be said for the more traditional spacetime where shell collapse results in an eternal Schwarzschild BH. Would the iteration process ever provide the drastic change in causal structure needed to account for the BH disappearance at the end of evaporation? If not, then performing QFT calculations in a singular non-evaporating background, while useful for gaining general intuition, is useless for obtaining a complete description of the BH evaporation process. To get legitimate candidates for a self-consistent description, QFT should be performed in spacetimes like~Fig.~\ref{fig:nonsingular-realistic}. Various efforts are making interesting progress in this direction, especially~\cite{Frolov:2016gwl,Frolov:2017rjz}, as discussed in previous sections. Moreover, a similar criticism can be applied to some of the firewall arguments:~quantum firewall states calculated in a vacuum spacetime are not self-consistent.

There is one more key point which has not been addressed, relating to the fact that the matter in Hawking radiation should be ``the same" matter that fell in, and, consequently, relating to self-consistency of the semiclassical BH solutions. In all standard calculations, the matter which falls in to form the BH, and the matter in which the Hawking radiation is present, are regarded as \textit{separate matter fields}. This is implicit when the background metric is assumed, in which case the infalling matter field is non-dynamical. A more conceptually accurate treatment would treat infalling matter as a true quantum field, with this being the only matter field in the problem. Then the true quantum in-state is not an in-vacuum, but a collapsing-star state. One still expects to find the Hawking radiation, but its source and backreaction effect would, presumably, be less mysterious. It may be the case that the standard calculation is appropriately regarded as a perturbation in this scenario. But the consequences of resolving this oversimplification have never been adequately settled.

Setting aside the discussion of its validity, assuming the non-future-directed-trajectory interpretation we have here espoused would lead to the following narrative of BH evolution: A star collapses beyond its Schwarzschild limit, forming a trapped region. Once the trapped region is formed, continued collapse is inevitable until quantum gravity takes over the dynamics, at which point a tiny core of Planck scale density is formed. The future boundary of the core acts, by some currently unknown mechanism associated with the core's extreme density, as a local space-and-time-dependent potential barrier, suppressing causal propagation of matter out of the core, and forcing matter propagation to be dominated by non-causal paths exiting the core through its spacelike outer surface. Thus, having effectively no other option, matter from the core slowly leaks (or ``tunnels") out on these non-causal paths, gradually depositing the BH mass at infinity in the form of a Hawking flux. Matter falls in, gets decomposed in the fires of quantum gravity, and, retaining unitarity, eventually comes back out. In terms of relativistic causal structure, in this picture there is no profound difference between a BH and other objects; black holes are unusual only in that much of their interior is empty and contains trapped surfaces, in that their matter is confined to a tiny volume where quantum gravity dominates, and for their low luminosity-to-mass ratio. In this picture, black holes are not that weird.

We are not the first to propose a picture like the one presented throughout this section. As far back as the original discovery of the Hawking effect, similar ideas were invoked in \cite{Hartle:1976tp}, and by various comments of~\cite{Hawking:1974sw,Hawking:1976ra} (though the particulars there are rather hazy, especially regarding different types of energy), as well as many others. It has not previously, however, been taken seriously as a semi-local physical description in the context of self-consistent evaporating nonsingular models. These models have the advantage that matter tunneling out has somewhere to come from, with an energy-conserving backreaction.

Whether the above is a useful or accurate story is undecided. It does, at least, seem qualitatively aligned with both the DFU stress tensor and the tunneling picture of Parikh and Wilczek~\cite{Parikh:1999mf} (the key step of which is a tunneling event in which some internal matter crosses the horizon), as well as with~\cite{Hawking:1974sw,Hawking:1976ra,Hartle:1976tp} as previously noted, and doesn't seem to raise any major philosophical issues. In~any case, the most appealing aspect of this description is surely its simplicity.

\section{XI. \ \ Concluding Remarks}

Spacetime diagrams invariably have a profound influence on our thinking about relativistic systems, and especially black holes (BHs). The goal of this article is to disrupt an unfortunate status quo: the use of diagrams not tied to any particular spacetime model. Ambiguous hand-drawn diagrams (most perniciously those attempting to depict BH formation and evaporation in a single picture) too often reflect the biases of the artist, and result in misleading intuitions --- for example that Hawking radiation emanates from (just outside) a null event horizon, or that external observers take an infinite proper time to see infalling matter fall in. False intuitions like these can lead to incorrect or circular reasoning about subtle questions.  To clarify these issues, we have argued for the use of well-defined models in which verifiable claims can be made. Our ``shell model," as it has been dubbed above, presents an attempt at a simple and minimal concrete model for BH formation and evaporation, which seems to capture most generic aspects of the problem, and for which Penrose diagrams can be explicitly obtained. Based on the results, we have argued for an improved definition of the term ``black hole," and proposed a more straightforward interpretation of the mechanism of Hawking radiation. While we make no claim of the absolute veracity of our shell model, we do hope it brings to light some new questions about the BH evaporation process, and stimulates a more concrete and physically grounded discussion.

\vspace{10pt}
\begin{acknowledgments}
This research was supported by the Foundational Questions Institute (FQXi.org), of which AA is Associate Director, and by the Faggin Presidential Chair Fund.
\end{acknowledgments}




\bibliography{form_evap}



\clearpage

 \section{Appendix \\[6pt] Matter content of Shell Models: General Case}

 Numerically computing Penrose diagrams allows the distribution and flow of matter to be quantitatively visualized in the diagram, assuming Einstein's equation $G_{\mu \nu}=8 \pi T_{\mu \nu}$. There are two contributions to the matter content: matter associated with the shells, and matter associated with the quasistatic equilibrium solutions.
 
 \subsection{Quasistatic Contribution}
 
 First we consider the matter associated with the equilibrium solutions. For our purposes, all such solutions take a metric of the form
 \begin{equation}
 \label{eqn:sss-metric}
     ds^2=-f(r)\,dt^2+f(r)^{-1}\,dr^2+r^2\,d\Omega^2 \, ,
 \end{equation}
 where $f(r)$ is an arbitrary function called the \textit{metric function}. Metrics of this type can be either singular or nonsingular at the origin. A number of equivalent conditions for singularity are given by \cite[ app. E]{Schindler:2018wbx}; here it suffices to say that the metric is \textit{nonsingular} whenever $f(r)=1+\mathcal{O}(r^2)$ as $r \to 0$, a condition which implies  finiteness of curvature scalars, geodesic completeness, and existence of a Cartesian metric, in a neighborhood of the origin \cite{Schindler:2018wbx}. Although classical theorems predict singularity formation in gravitational collapse \cite{Hawking:1973uf}, nonsingular solutions are thought to arise in effective semiclassical approximations if quantum gravitational effects regulate curvature at the Planck scale. Our method applies to both singular and nonsingular models; nonsingular models have the advantage that all matter is made explicit in the stress tensor, whereas singular solutions contain a matter contribution hidden in the singularity.

 \paragraph{Curvature.}
 A detailed anaylsis of the matter content for metrics of the form (\ref{eqn:sss-metric}) was carried out in \cite{Schindler:2018wbx}. The following is a summary of those results. To expedite the analysis, it is best to define a \textit{mass function} $m(r)$ by
 \begin{equation}
     \label{eqn:mass-fn}
     f(r) = 1 - 2 \, m(r)/r ,
 \end{equation}
 and define an orthonormal basis $\ehat_a$ by 
\begin{equation}
    \begin{array}{l}
    \ehat_0 = \left\lbrace
        \begin{array}{ll}
            \sqrt{f(r)^{-1}} \; \pp_t  &, f(r)>0, \\
            \sqrt{-f(r)}     \; \pp_r  &, f(r)<0,
        \end{array} \right.
    \\[13pt]
    \ehat_1 = \left\lbrace
        \begin{array}{ll}
            \sqrt{-f(r)}     \; \pp_r  &, f(r)>0, \\
            \sqrt{f(r)^{-1}} \; \pp_t  &, f(r)<0,
        \end{array} \right.
    \\[15pt]
    \ehat_2 = r^{-1} \; \pp_{\theta},
    \\[6pt]
    \ehat_3 = (r \sin \theta)^{-1} \; \pp_{\phi} \, .
\end{array}
\end{equation}
In this basis $\ehat_0$ is always timelike. Both $\ehat_0$ and $\ehat_1$ can be continuously extended across the horizons where \mbox{$f(r)=0$}, but the full basis cannot, since the extensions of $\ehat_0$ and $\ehat_1$ would coincide at the horizon. In this $\ehat_a$ basis, the Einstein tensor is diagonalized, with components
\begin{equation}
    \label{eqn:Gab}
    {G^{a}}_{b} = 8 \pi \; \diag(-\rho, -\rho,p_\Omega,p_\Omega),
\end{equation}
where
\begin{equation}
\label{eqn:density-pressure}
    \rho = \frac{m'(r)}{4\pi r^2},
    \qquad
    p_\Omega = - \frac{m''(r)}{8\pi r}.
\end{equation}
Physically, this amounts to a proper density $\rho$, a transverse pressure $p_t = -\rho$, and an angular pressure $p_{\Omega}$. The common curvature scalars follow, as 
\begin{equation}
    \begin{array}{l}
         K_0 \equiv R = 16 \pi \, (\rho - p_\Omega),
        \\[2pt]
        K_1 \equiv R_{ab}R^{ab} = 128 \pi^2 \, (\rho^2 + p_\Omega^2),
        \\[2pt]
        K_2 \equiv C_{abcd} C^{abcd} =  12 \, \eta^2 / r^4 ,
        \\[2pt]
        K_3 \equiv R_{abcd} R^{abcd} = K_2 + 2 \, K_1 - (1/3) \, K_0^2 ,
    \end{array}
\end{equation}
with $\eta = 2 \, m(r)/r - 4 \, m'(r)/3 + r \, m''(r)/3 $. Contributions to the curvature from a singularity, if one exists, are included in $K_2$, and $f(r) \equiv 1$ if and only if $K_1=K_2=0$ everywhere. A complete specification of the Riemann and Weyl curvature components, in addition to Christoffel symbols, may be found in \cite{Schindler:2018wbx}.

\paragraph{Energy conditions.}
Nonsingular black hole solutions often violate classical energy conditions---this is one way to evade the singularity theorems \cite{Hawking:1973uf}. One approach to this situation is to take both the nonsingular metric and its energy condition violations seriously, assuming they provide useful insight about the physical mechanism of evaporation. Although classically unorthodox, this approach is appealing since quantum field theories are already well known to predict energy condition violations \cite{Martin-Moruno:2017exc}. Regardless of one's view on this matter, it is useful to keep track of where and by how much energy conditions are violated in a given solution. We concern ourselves here with the null (NEC), weak (WEC), and flux (FEC) energy conditions, defined by \cite{Martin-Moruno:2017exc}
\begin{equation}
    \label{eqn:ecs}
    \begin{array}{cl}
        (\textrm{NEC}) & G_{ab} \, k^a k^b \geq 0 \;\; \textrm{for all null } k^a, \\
        (\textrm{WEC}) & G_{ab} \, t^a t^b \geq 0 \;\; \textrm{for all timelike } t^a, \\
        (\textrm{FEC}) & -{G^{a}}_{b} \, t^b \;\; \textrm{causal for all timelike } t^a.
    \end{array}
\end{equation}
The WEC ensures timelike observers measure locally positive mass density, with the NEC as its null limit; the FEC ensures no timelike observer measures a spacelike energy flux. Given the diagonalized Einstein tensor (\ref{eqn:Gab}), these reduce to the simple inequalities \cite{Martin-Moruno:2017exc}
\begin{equation}
    \label{eqn:ec-ineqs}
    \begin{array}{cl}
        (\textrm{NEC}) & \rho + p_{\Omega} \geq 0, \\
        (\textrm{WEC}) & \textrm{NEC plus } \rho \geq 0, \\
        (\textrm{FEC}) & \rho^2-p_{\Omega}^2 \geq 0.
    \end{array}
\end{equation}
The degree to which these conditions are violated is quantified by the functions
\begin{equation}
    \begin{array}{rcl}
        \nec &=& - \min(\rho + p_{\Omega}, 0), \\
        \wec &=& - \min(\rho, 0), \\
        \fec &=& - \min(\rho^2-p_{\Omega}^2, 0),
    \end{array}
\end{equation}
called the \textit{energy condition violation functions}. These functions are the obvious extensions of the above inequalities, but their physical status as quantifiers requires some clarification. They are justified as follows. Consider observers associated with a null velocity $k_a$ and a normalized timelike velocity $t_a$, and the flux vector $F^a = -{G^{a}}_{b} \, t^b$ relative to $t^b$ (note that the FEC is equivalent to $-F_a F^a \geq 0$). One then finds that
\begin{equation}
    \begin{array}{rcl}
        G_{ab} \, k^a k^b &=& ((k_2)^2+(k_3)^2) \; (\rho + p_{\Omega}), \\
        G_{ab} \, t^a t^b &=& ((t_2)^2+(t_3)^2) \; (\rho + p_{\Omega}) + \rho, \\
        -F_a F^a &=& ((t_2)^2+(t_3)^2) \; (\rho^2 - p_{\Omega}^2) + \rho^2 .
    \end{array}
\end{equation}
 Thus, for a given observer, the functions $\nec$ and $\wec$ quantify the amount of measured negative mass density, and the function $\fec$ quantifies the spacelike-ness of the energy flux. Interestingly, when $\rho>0$, only observers with large angular momentum will observe strong energy condition violations. This quantification scheme is in line with the standard definitions for semiclassical energy conditions \cite{Martin-Moruno:2017exc}, which usually amount to enforcing a small positive bound on our energy condition violation functions.

\paragraph{Trapped surfaces, horizons.}
An important characteristic of black hole spacetimes is the existence of closed trapped surfaces (see~\cite{Senovilla:2011fk} for a useful review); their existence is associated with the trademark ``inevitability" of black hole collapse. Naively, in applying the theory of trapped surfaces to study black holes, one basically wants to identify the region containing trapped surfaces and determine its boundary. The boundary of the trapped region in spacetime is sometimes called a ``trapping horizon," while the boundary of the region of trapped surfaces contained entirely within a spatial slice is often called an ``apparent horizon." These naive definitions capture the right essential spirit, but fall somewhat short at a technical level, mainly due to the possibility of strangely shaped trapped surfaces and the associated issue of ``clairvoyance"~\cite{Senovilla:2011fk}. Fortunately, an illuminating and thorough discussion of these issues, and their application to black hole spacetimes, has been carried out by Bengtsson and Senovilla~\cite{Bengtsson:2010tj}. They have determined that in spherical symmetry the trapped region, which is unreasonably global due to clairvoyance, contains an essential and physically relevant spherical subregion called ``the core of the trapped region," which we refer to here as the \textit{trapping nucleus} (to distinguish it from the entirely unrelated ``matter core" in the Hayward metric). The trapping nucleus is defined as a minimal region which, if removed from the spacetime, eliminates all trapped surfaces. That is, any trapped surfaces in spacetime can be blamed on the nucleus, even if they extend outside it. This definition is vindicated by the fact that the boundary of the trapping nucleus in spherical symmetry is the unique spherically symmetric marginally trapped tube, i.e., the apparent horizon at $r=2m$. Outside spherical symmetry there are many non-equivalent trapping nuclei for a trapped region, but it is conjectured that there is a unique trapping nucleus whose boundary is a marginally trapped tube~\cite{Bengtsson:2010tj}. It is on this special trapping nucleus, whose boundary we call \textit{the} apparent horizon, that we focus our attention in defining the black hole. Bengtsson and Senovilla have argued that the trapping nucleus and its boundary could provide the best trapped-surfaces definition of a black hole and horizon.

\paragraph{Trapped spheres.}
Following~\cite{Bengtsson:2010tj}, an arbitrary spherically symmetric spacetime can be expressed in the Eddington-Finklestein form
\begin{equation}
    ds^2 = -e^{2\beta} \, (1-2m/r) \, dv^2 + 2e^{\beta} \, dv \, dr + r^2 \, d\Omega^2,
\end{equation}
where $m=m(v,r)$ and $\beta=\beta(v,r)$. Expressed this way, the trapping nucleus (see above) is the set $r<2m$ and its boundary is $r=2m$. In our shell model, these correspond to the set $f(r)<0$ and its boundary $f(r)=0$. This nucleus is precisely the region in which spheres about the origin of spherical symmetry are trapped~\cite{Schindler:2018wbx,Bengtsson:2010tj}. To avoid unfamiliar terminology, we refer to this region (which is both the trapping nucleus and the set of points where the sphere $(t_0,r_0,\Omega)$ is a closed trapped surface) in the main text as the \textit{trapped spheres region}.

\paragraph{Trapping and horizons: summary.}
The trapped region surrounding a spherical black hole has an indispensable interior subset called the trapping nucleus, whose boundary is the apparent horizon. In our shell model, the trapping nucleus exactly coincides with the trapped spheres region $f(r)<0$, so that the apparent horizon is exactly at $f(r)=0$. For metrics with a nearly-Schwarzschild exterior of mass $m_0$, this implies the apparent horizon is almost exactly at $r=2m_0$.

\subsection{Shell Contribution}

The junction hypersurface connecting two properly matched piecewise-defined regions of spacetime in general corresponds to a thin shell of matter. This is described technically by a distributional contribution to the stress tensor of the joint spacetime. For the purposes of this article, we consider shells arising from the junction of spacetimes of the form (\ref{eqn:sss-metric}) along radial null hypersurfaces (excluding the horizon-matching case where $f(r)=0$ everywhere). The stress tensor for this setup was calculated in \cite{Schindler:2018wbx} by application of the null shell formalism of Barrabes and Israel \cite{Barrabes:1991ng}. The result is most concisely described using local Eddington-Finklestein coordinates in a neighborhood of the junction shell, defined as follows. 
Consider a local patch $M_0$ of the joint spacetime, which is separated into a past region $M_{-}$ and future region $M_{+}$ by the null junction hypersurface $\Sigma$, with metric functions $f_{\pm}(r)$ in the two regions. As shown in \cite{Schindler:2018wbx}, it is possible to choose a joint coordinate system $(w,r,\Omega)$ on $M_0$, such that the shell $\Sigma$ is defined by the level set $w=0$, and such that the metric is 
\begin{equation}
    ds^2 = - f(r) \, dw^2 - 2 \epsilon \, dw \, dr + r^2 \, d\Omega^2,
\end{equation}
where the parameter
\begin{equation}
    \epsilon = \left\lbrace
    \begin{array}{l}
        \textrm{$-1$ if $\pp_r$ is past-directed}, \\
        \textrm{$+1$ if $\pp_r$ is future-directed},
    \end{array} \right.
\end{equation}
is a constant indicating whether the shell $\Sigma$ is ingoing ($\epsilon=-1$) or outgoing ($\epsilon=+1$) towards the future, and the metric function
\begin{equation}
    f(r) = \left\lbrace
    \begin{array}{ll}
        f_{-}(r), &  w<0, \\
        f_{+}(r), &  w>0,
    \end{array} \right.
\end{equation}
is defined piecewise on $M_{\pm}$. Let
\begin{equation}
    n^{\mu} = \epsilon \, (\pp_r)^{\mu}
\end{equation}
be a future-directed null vector both normal to and tangential to $\Sigma$, let the mass functions $m_{\pm}(r)$ be defined according to (\ref{eqn:mass-fn}), and define the \textit{mass jump} $[m(r)]$ by
\begin{equation}
    [m(r)] =  m_{+}(r) - m_{-}(r).
\end{equation}
With this setup, the distributional component of the stress tensor on the shell obtains the simple expression 
\begin{equation}
    \label{eqn:Tuv-shell}
    T^{\mu\nu}_{\Sigma} =  \sigma \, n^{\mu} n^{\nu} \; \delta(w) ,
\end{equation}
where
\begin{equation}
   \sigma = (-\epsilon) \, \frac{[m(r)]}{4\pi r^2} \, .
\end{equation}
The coefficient $\sigma$ may be thought of as the surface energy density of the shell, up to an arbitrary normalization factor associated with the null vector $n^a$. The sign of $\sigma$ is physically meaningful: timelike observers measure a positive energy density at the shell if and only if $\sigma>0$. It is therefore sensible to say that shells with $\sigma<0$ have \textit{negative mass}, while shells with $\sigma>0$ have \textit{positive mass}. The sign of $\sigma$ is a local property, and in principle (in physically unusual cases) a single shell may have positive and negative mass at different points.

\paragraph{Energy conditions.} The shell stress tensor (\ref{eqn:Tuv-shell}) is easily analyzed in terms of the energy conditions (\ref{eqn:ecs}) above. For an arbitrary causal vector $u^a$ such that $u_a u^a \leq 0$,
\begin{equation}
    \begin{array}{rcl}
        T_{\mu\nu} u^{\mu} u^{\nu} & = & \sigma \; (- n_{\nu} u^{\nu})^2 \; \delta(w) , \\[2pt]
        -{T^{\mu}}_{\nu} u^{\nu} & = &  \sigma \; (-n_{\nu} u^{\nu}) \;  n^{\mu} \; \delta(w) . \\[1pt]
    \end{array}
\end{equation}
Thus the WEC and NEC are violated if and only if $\sigma < 0$. Since the flux vector $F^\mu = -{T^{\mu}}_{\nu} u^{\nu}$ is always null, the FEC is not particularly meaningful in this context. It is worth noting, however, that if $u^\mu$ is future-directed (implying $n_\mu u^\mu \leq 0$), then $F^\mu$ is future-directed if and only if $\sigma > 0$. These considerations support the above notion that $\sigma >0$ corresponds to normal matter, while $\sigma <0$ corresponds to exotic matter.

\subsection{Energy Conservation and DTR}

Local energy conservation, of the form $\del_\mu T^{\mu \nu} = 0$, is automatically guaranteed at points where the metric is smooth, and along properly matched shell junctions~\cite{Barrabes:1991ng}. At points where shells collide, for example at Hawking radiation nucleation points in the above model, energy conservation must be independently verified by checking an equality called the DTR (Dray-'t Hooft-Redmount) relation \cite{Barrabes:1991ng}. In case of two radial null shells colliding at radius $r_0$, separating spacetime into four regions (each of the form (\ref{eqn:sss-metric})) labeled $A,C,B,D$ clockwise from noon (see Fig.~\ref{fig:shell-collision}), the DTR relation reads \cite{Schindler:2018wbx}
\begin{equation}
    f_A(r_0) f_B(r_0) - f_C(r_0) f_D(r_0) = 0 .
\end{equation}
To quantify violations of energy conservation, we therefore define the \textit{DTR violation function}
\begin{equation}
    \dtr = |f_A(r_0) f_B(r_0) - f_C(r_0) f_D(r_0)|,
\end{equation}
so that energy conservation is equivalent to $\dtr=0$. 

Applied to the Hawking radiation nucleation points, at which $f_B(r)=f_C(r)=f_D(r)$, this yields
\begin{equation}
    \dtr = |f_A(r_0) - f_B(r_0)| \, |f_B(r_0)|.
\end{equation}
Assuming that the shells carry a finite amount of mass, and that the nucleation radius $r_0$ is a finite radial distance $\levap$ outside the horizon where $f_B(r)=0$, it follows that both terms above are finite, so
\begin{equation}
    \dtr > 0,
\end{equation}
and energy conservation is violated at nucleation points.

Although the DTR relation is not satisfied at the Hawking radiation nucleation points, the violation is arbitrarily small in the physically relevant limits of the model. In particular, in the limit $\levap \to 0$ in which nucleation points approach the horizon radius, the factor $|f_B(r_0)| \to 0$ while the other remains finite, so that 
\begin{equation}
    \dtr \to 0.
\end{equation}
Physically, it is likely that this limit should be taken only down to the Planck scale, so that $\levap \approx \lpl$, in which case one would interpret the violation of energy conservation to represent a small quantum fluctuation. Since energy conservation is restored in the physically relevant limits of the model, the model remains a useful approximation to physically realistic spacetimes.






\end{document}